\documentclass[preprint,12pt]{elsarticle}

\usepackage{lineno}
\usepackage{siunitx}
\usepackage{graphicx}%
\usepackage{multirow}%
\usepackage{physics}
\usepackage{amsmath}%
\usepackage{amssymb}%
\usepackage{amsfonts}%
\usepackage{amsthm}%
\usepackage{mathrsfs}%
\usepackage[title]{appendix}%
\usepackage{xcolor}%
\usepackage{textcomp}%
\usepackage{manyfoot}%
\usepackage{booktabs}%
\usepackage{algorithm}%
\usepackage{algorithmicx}%
\usepackage{algpseudocode}%
\usepackage{listings}%
\usepackage[version=3,arrows=pgf-filled ]{ mhchem}
\usepackage{float}
\usepackage{siunitx}
\usepackage[nice]{nicefrac}
\usepackage{gensymb}
\usepackage{wasysym}
\usepackage{todonotes}
\usepackage{makecell}
\usepackage{lmodern}
\usepackage{mathrsfs}
\usepackage{wasysym}
\usepackage{adjustbox} 
\usepackage{url}
\newcommand{\bareFrict}{\gamma} 
\newcommand{\velFluid}{\vb{v}_\text{fluid}}
\newcommand{\velBact}{\vb{v}_\text{bact}} 
\newcommand{\beadPos}[1]{\vb{r}_{#1}}
\usepackage[colorlinks=true,linkcolor=blue,citecolor=blue,urlcolor=blue]{hyperref}
\usepackage{cleveref}%% The amsthm package provides extended theorem environments
\usepackage{subcaption}
\journal{Advances in Water Resources}

\begin{document}

\begin{frontmatter}



\title{Intermittent flow paths in biofilms grown in a microfluidic channel}

\author[1]{Kerem Bozkurt*} %% Author name
\author[2]{Christoph Lohrmann}
\author[1,3]{Felix Weinhardt}
\author[1]{Daniel Hanke}
\author[1]{Raphael Hopp}
\author[4]{Robin Gerlach}
\author[2]{Christian Holm}
\author[1]{Holger Class}
\affiliation[1]{organization={Institute for Modelling Hydraulic and Environmental Systems; University of Stuttgart},%Department and Organization
            addressline={Pfaffenwaldring 61}, 
            city={Stuttgart},
            postcode={70569}, 
            state={Baden Württemberg},
            country={Germany}}

\affiliation[2]{organization={Institute for Computational Physics, University of Stuttgart},%Department and Organization
            addressline={Allmandring 3}, 
            city={Stuttgart},
            postcode={70569}, 
            state={Baden Württemberg},
            country={Germany}}

\affiliation[3]{organization={Department Technical Biogeochemistry, Helmholtz Centre for Environmental Research},%Department and Organization
            addressline={Permoserstraße 15}, 
            city={Leipzig},
            postcode={04318}, 
            state={Baden Württemberg},
            country={Germany}}
            
\affiliation[4]{organization={Department of Chemical and Biological Engineering and Center for BiofIlm Engineering, Montana State University},%Department and Organization
            city={Bozeman},
            postcode={MT 59717}, 
            state={Montana},
            country={USA}}
%% Abstract
\begin{abstract}
Biofilms exposed to flow experience shear stress, which leads to a competitive interaction between the growth and development of a biofilm and shearing. In this study, \textit{Pseudonomas fluorescene} biofilm was grown in a microfluidic channel and exposed to forced flow of an aqueous solution of variable velocity. It can be observed that under certain conditions preferential flow paths form with a dynamic, but quasi-steady state interaction of growth, detachment, and re-attachment. 
We find that the regimes for preferential flow path development are determined by nutrient availability and the ratio of shear stress versus the biofilm's ability to resist shear forces.
The intermittent regime of flow paths is mainly driven by the supply with  nutrients, which we confirm by comparison with a numerical model based on coarse-grained molecular dynamics and Lattice Boltzmann hydrodynamics.
\end{abstract}

\begin{graphicalabstract}

\begin{figure}[H]
    \centering
    \includegraphics[width=\linewidth]{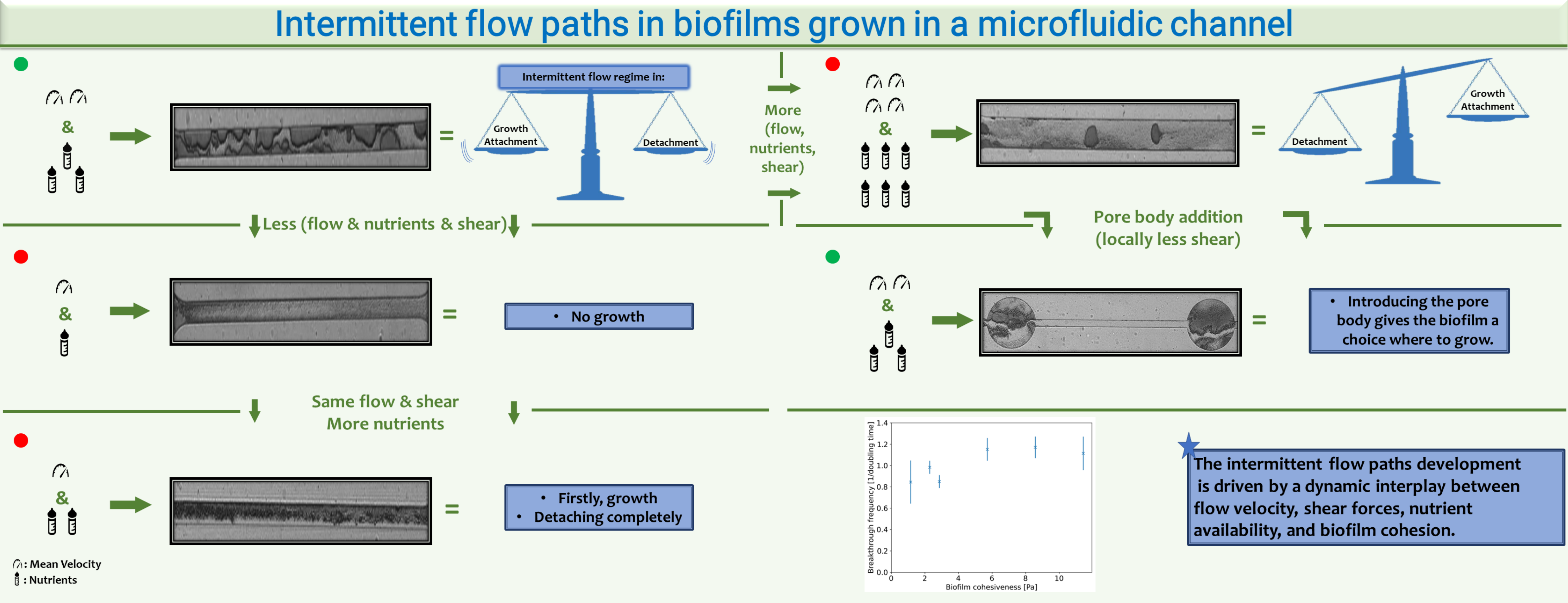}
\end{figure}

\end{graphicalabstract}

\begin{highlights}
 \item Biofilm growth with preferential flow paths depends on shear and nutrient conditions.
    \item Biofilms can develop by adapting to varying growth conditions and environments.
    \item Biofilm cohesiveness impacts the development of flow paths but it does not effect the frequency of the detachment events significantly.
    \item The frequency of detachment events is influenced by the flow velocity and the availability of nutrients.
\end{highlights}

%% Keywords

\end{frontmatter}



%% Use \subsection commands to start a subsection.
\section{Introduction}
\label{sec:intro}
%% Labels are used to cross-reference an item using \ref command.

Fluid flow in porous media plays a fundamental role in many natural and industrial processes, such as in reservoir rocks, groundwater aquifers, and water filtration systems. The flow within groundwater aquifers is primarily determined by porosity, which is the ratio of void volume to total volume \cite{Sheer2010}. The porosity of a porous medium can evolve due to various processes, e.g., due to the formation of microbial biofilms which are communities of microorganisms \cite{Cunningham1991, Hommel2018}. Microorganisms in biofilms are encased in a matrix of extracellular polymeric substances, such as exopolysaccharides and proteins, which shield the bacteria from external factors and can affect the porosity, permeability and transformation processes in porous media \cite{Stoodley2004, Davey2000, Characklis1981, Costerton1987}. For instance, biofilms can enhance reaction rates in biomineralization \cite{Weinhardt2021}, filter or degrade harmful substances in bioremediation, and act as biobarriers to prevent \ce{CO2} leakage \cite{Ebigbo2010}. Additionally, the biofilm growth mode protects bacteria from environmental stresses and substances like disinfectants and antibiotics. In industrial settings biofilms can cause significant damage to materials as well as energy losses and blockages in condenser tubes and heat exchange tubes \cite{Sandholm2009,Characklis1981rep}. As a result, the successful development of many technologies requires an understanding of how biofilms develop as they interact with flow in porous media. 

Biofilm growth in porous media often leads to pore clogging, which alters both porosity and hydraulic conductivity. However, clogged pores can reopen when the hydraulic forces exerted on the biofilm are strong enough to shear it off \cite{Weinhardt2021, Cunningham1991, Hommel2018}. Therefore, biofilms must withstand flow and shear forces throughout all stages of their formation, which include initial reversible attachment, irreversible attachment, maturation, and finally, dispersion \cite{Sauer2002,Sharma2023, Wheeler2019, lohrmann23a}.  In other words, while fluid flow can increase the transport of nutrients to the microbes within biofilms and increase growth, the developing shear stress can also result in increased detachment of microbes \cite{Wheeler2019, Hommel2015, lohrmann23a, Kurz2022, Allen2018}. 

At low shear, biofilms become more porous and permeable while at high shear, they are generally more cohesive, denser and stiffer \cite{Chen2005, Aggarwal2015, Stoodley2002, Klapper2002, Fanesi2021, Characklis1981}. Despite their high cohesion and adhesion, biofilm detachment can still occur under high shear stresses. If the local stress exceeds cohesive forces, biofilm detachment is observed \cite{Ochoa2007}. When exposed to high shear stresses, biofilms tend to form filamentous structures. Consequently, deformations such as necking can occur, creating weak points that may break internally, leading to cohesive failure and subsequent detachment \cite{Stoodley2002}. A net accumulation of biomass will be observed in total when biomass adsorption, cell growth, attachment, and filtration exceed biomass desorption and detachment \cite{Cunningham1991}.

Under certain conditions, biofilms can detach from surfaces, reattach, and re-grow biofilms. Consequently,  preferential flow paths which allow fluid transport can develop. These flow paths may periodically become narrow or even close, only to reopen and allow flow again. The periodicity of these events leads to spatial and temporal rearrangements, which is referred to as an intermittent regime \cite{Kurz2022}. In this intermittent regime, the frequency of flow paths is a measure that expresses how often the flow paths open, close, and reorganize over time.
Understanding how these flow paths and consequently their intermittency evolve is crucial as it impacts the transport of fluids e.g. nutrients or contaminants, playing an essential role in managing bioremediation, wastewater treatment, and microbial behavior in porous media. Additionally, it enables the development of strategies to enhance biofilm-based processes and prevent undesired biofilm accumulation.

Thus, experimental and numerical studies arose to investigate biofilm growth in various media in the presence of flow, also focusing on the mechanism of preferential flow paths \cite{Zhou2020,Kurz2022,Bottero2013,Stewart2001,Sharp2025, Durham2012, Gaol2021}. \citet{Stewart2001} as well as \citet{Gaol2021} found that the dynamics of bacterial growth lead to the formation of preferential flow paths in investigated biomass plugging porous media. Expanding upon this, \citet{Sharp2025} also conducted experiments roughly 4 cm x 8 cm flat panel reactors with 1 mm porous media elements that thicker biofilms exhibit more flow paths, whereas thinner biofilms tend to distribute the flow more homogeneously throughout the biofilm-affected porous media. \citet{Zhou2020} investigated the effect of pore structure on biofilm distribution and growth. They observed a relationship between the distribution of the biofilm within the flow cells and the development of flow paths. They found that flow cells with angular shaped elements have more flow paths than flow cells with spherical-shaped elements and that because the flow paths provide more nutrients (including potentially oxygen), biofilm tends to accumulate faster in these places. \citet{Kurz2022} also observed the relationship between biofilm growth and flow paths. They stated that the closure of flow paths is due to microbial growth, while their reopening is due to shear forces. They observed an increase in shear rate as flow paths narrow, and shear forces decrease when the flow path widens again. \citet{Bottero2013} conducted a modeling study on the relationship between biofilm growth and flow paths, considering decay and lysis. In one case, they excluded decay and lysis and observed detachment due to shear forces following biofilm growth.  However, because of the blockage in isolated areas, the flow always followed the same path. In scenarios with decay only, microbial cells far from open paths decayed, but detachment did not occur without shear stress. When lysis was added to the simulation, microbial cells in blocked areas broke down, causing detachment and forming new flow paths. These summarize previous findings \cite{Zhou2020,Kurz2022,Bottero2013,Stewart2001,Sharp2025, Durham2012, Gaol2021} collectively highlight the critical role of microbial growth, shear forces, and pore structure in shaping flow paths and influencing biofilm dynamics in porous media. Building on the lessons from these studies, we investigate how shear forces and nutrient availability influence the structural and rheological properties of biofilms, thereby modulating the flow paths. 

In order to further explore these effects, we compare and present here the results of a numerical model coupled to Lattice-Boltzmann hydrodynamics and experiments on biofilm growth under flow conditions in microfluidic channels. In our experiments, we grew biofilms at varying flow rates and observed how the biofilm responded to various shear stresses. We employed designs that were either just a basic flow channel or a flow channel with additional pore bodies. This enabled us to clearly observe biofilm growth in intermittently opening and closing flow paths. Furthermore, we extended our previous numerical model \cite{lohrmann23a} for biofilm growth and compared it with the experimental results under similar conditions. 

In the following, we provide an overview of our experimental setup, the cultivation conditions, the production of the microfluidic channels, the experimental procedures, and the development of simulations. Then, using both the numerical model and the experimental findings, we present and discuss representative results. In the final section, we provide a summary of our research and some key findings. Our work indicates that the development of preferential flow paths is dependent on the biofilm's ability to resist shear forces.

\section{Methods}\label{sec:methods}

\subsection{Experimental}\label{subsec2}

\subsubsection{Materials}\label{subsubsec2}
\subsection*{Bacterial culture}
The bacterial strain used in this study is \textit{Pseudomonas fluorescens} (DSM 50000 106). \textit{P. fluorescens} is a rod-shaped, Gram-negative, motile, aerobic species. Pseudomonas occur frequently in both aquatic and soil environments and have been described to produce biofilms, promote plant growth and be antibiotic resistant \cite{Naik2023}. The optimal growth conditions are temperatures between 25 °C and 30 °C, with a lower limit of 4 °C and an upper limit of 37 °C \cite{Scales2014}.

\textit{Pseudomonas fluorescens} was stored on agar plates at 8°C prior to the experiments. To conduct the experiment, the bacteria were prepared in a liquid culture using sterile deionized water and 8\,\nicefrac{g}{l} of nutrient broth powder.
One colony of bacteria were transferred into 250 mL flask containing 60~ml of nutrient broth using an inoculation loop. They were incubated by shaking at 30°C at 180 rpm overnight.
The optical density ($\text{OD}_{600 nm}$) was measured using a spectrophotometer with a path length of 1 cm before the experiment.
In the end, the bacterial culture was diluted to an $\text{OD}_{600 nm}$ of 1.

\subsection*{Microfluidics}
\begin{figure}[H]
    \centering
    \includegraphics[width=\linewidth]{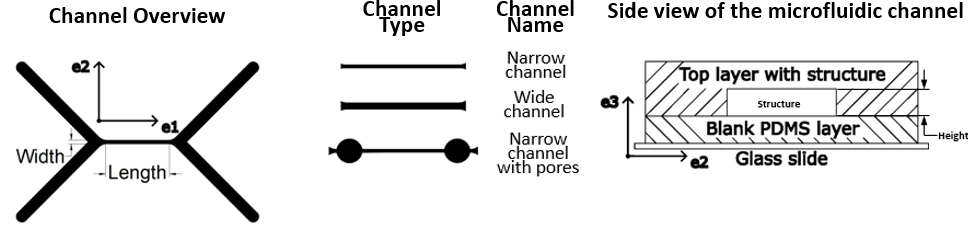}
    \caption{Flow channel overview, channel types and their names, and side view of the flow channel}
    \label{fig:celloverview}
\end{figure}

The PDMS (Polydimethylsiloxane) based microfluidic channels were fabricated using the general workflow of soft lithography \citep{XiaWhitesides1998, Karadimitriouetal2013}.
The designs as shown in Figure~\ref{fig:celloverview} were created with AutoCAD\textsuperscript{©} and printed on A4 transparency masks. These masks were used in optical lithography with SU8-3050 photoresist to create spin-coated silicon wafers. These wafers were used to produce the actual microfluidic channels using PDMS \cite{Weinhardt2021}.
Dow Corning SYLGARD© 184 Silicone Elastomer base and its curing agent were mixed at a ratio of 1:10 \cite{Weinhardt2021, Nikos2012}
After degassing them in a vacuum chamber, they were poured into the two petri dishes, one with wafer and one without. Thereafter, both Petri dishes were put into the oven for 2h 10min. After cooling to room temperature for 10 minutes, the PDMS was removed from the dishes, and 0.5mm holes were punched into the gels to create inlets and outlets.
A corona treater was used to bond both sides together. After cutting out the microfluidic channel, connecting the tubing, and repeating the corona-treatment process with a glass slide to stick to the bottom side, the channel was ready for use. A side view of the channel layers can be seen in Figure~\ref{fig:celloverview}.

Figure \ref{fig:celloverview} shows a top down view of the channel created in AutoCAD\textsuperscript{©}. The lengths (4.1 mm) and depths (0.035 mm) of all flow channels are the same. The difference between the first two flow channels shown in Figure \ref{fig:celloverview} differs in terms of their channel widths. The width of the narrow channel is 0.125 mm, while that of the wide channel is 0.25 mm. Therefore, in the following sections, we will refer to these channels by their widths \textit{(narrow channels, wide channels)}. The third channel also has a width of 0.125 mm but additionally contains two pore bodies. We will refer to this as the \textit{narrow channel with pores}.
 in Table \ref{table: cell dimensions}

\begin{table}[htbp]
    \centering
        \caption{Dimensions of the used microfluidic channel}
    \resizebox{\textwidth}{!}{%
    \begin{tabular}{c c c c c}
        \hline
        \makecell{\textbf{Flow channel}} & \makecell{\textbf{Channel width,} \\ \textit{w}, in mm} & \makecell{\textbf{Channel height,} \\ \textit{h}, in mm} & \makecell{\textbf{Channel length,} \\ \textit{l}, in mm} & \makecell{\textbf{Pore radius,} \\ \textit{in mm}} \\
        \hline
        Narrow channel & 0.125 & 0.035 & 4.1 &  \\
        Wide channel & 0.25 & 0.035 & 4.1 &  \\
        Narrow channel with pores & 0.125 & 0.035 & 4.1 & 0.5 \\
        \hline
    \end{tabular}%
    }
    \label{table: cell dimensions}
\end{table}

\subsubsection{Experimental Setup and Procedure}\label{subsubsec3}

The setup consisted of a CETONI Base 120, which connected the syringe pumps and computer for process control. It included two CETONI neMESYS 100 N pumps: one operating a 2.5\,ml syringe for the bacterial liquid culture, and the other operating a 5\,ml syringe for the nutrient broth solution. CETONI QMixElements software was used to control flow, allowing for real time flow-rate changes and scripting the injection strategy prior to the injections. The syringes were connected to T-valves with PTFE-tubes (poly-tetra-fluoro-ethylene) which had an inner diameter of 0.5 mm and an outer diameter of 1.59 mm (1/16 in.). The channel was mounted onto a stage, which enabled image capturing using transmitted-light microscopy coupled with a camera. A detailed description of this optical setup can be found in \cite{Karadim2012}.

\begin{figure}[H]
    \centering
    \includegraphics[width=0.5\linewidth]{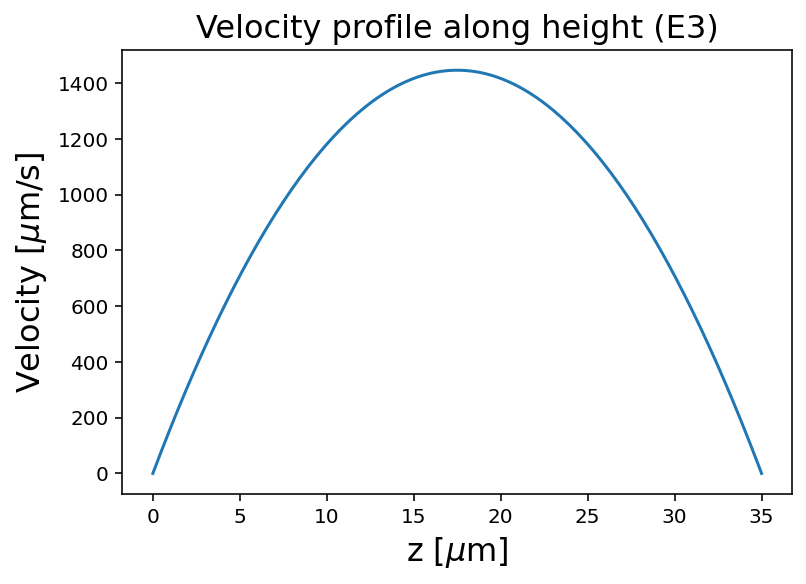}
    \caption{Velocity profile of E3 ($v_0 = 0.8 \, \text{mm/s}$ in narrow channel) along height (0.035 mm)}
    \label{fig:vpheight}
\end{figure}

\begin{figure}[H]
    \centering
    \includegraphics[width=0.5\linewidth]{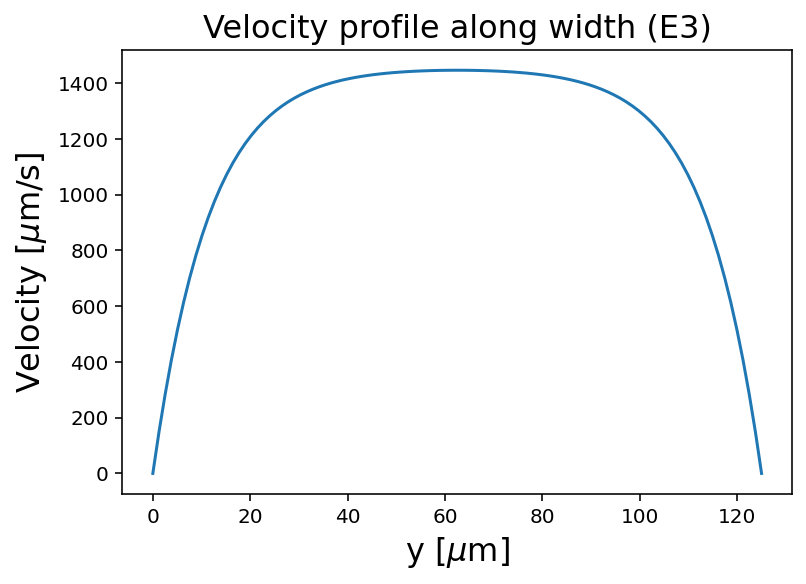}
    \caption{Velocity profile of E3 ($v_0 = 0.8 \, \text{mm/s}$ in narrow channel) along width (0.125 mm)}
    \label{fig:vpwidth}
\end{figure}

\begin{figure}[H]
    \centering
    \includegraphics[width=0.5\linewidth]{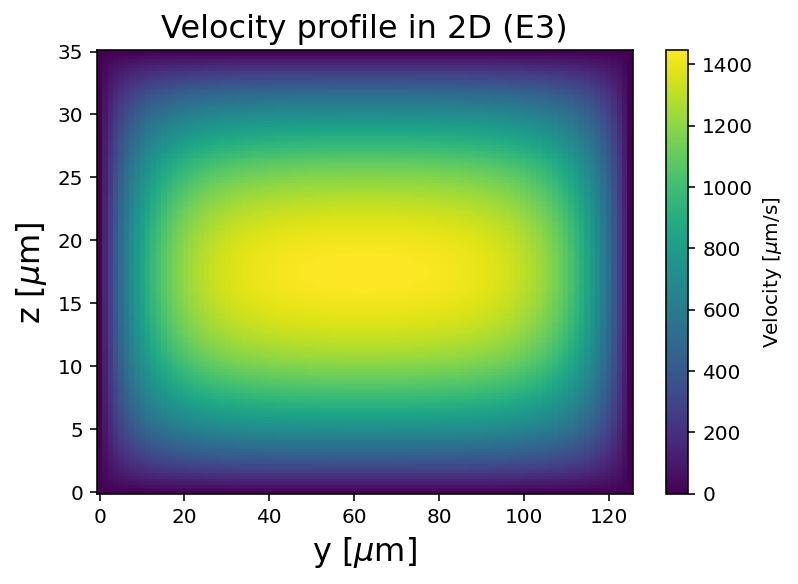}
    \caption{Velocity profile of E3 ($v_0 = 0.8 \, \text{mm/s}$ in narrow channel) in 2D}
    \label{fig:vp2D}
\end{figure}

\begin{figure}[H]
    \centering
    \includegraphics[width=0.5\linewidth]{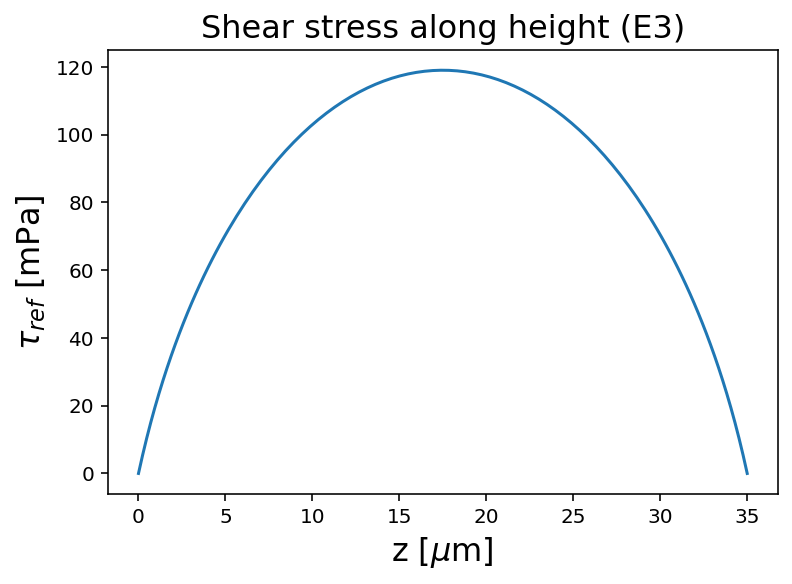}
    \caption{Shear stress of E3 ($v_0 = 0.8 \, \text{mm/s}$ in narrow channel) along height (0.035 mm)}
    \label{fig:shearheight}
\end{figure}

\begin{figure}[H]
    \centering
    \includegraphics[width=0.5\linewidth]{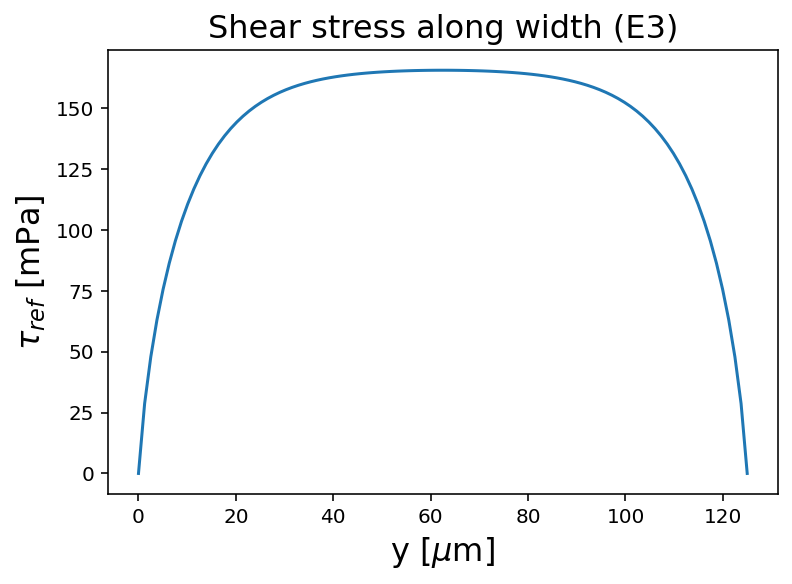}
    \caption{Shear stress of E3 ($v_0 = 0.8 \, \text{mm/s}$ in narrow channel) along width (0.125 mm)}
    \label{fig:shearwidth}
\end{figure}

The injection strategy consisted of two parts. The first part began with injecting 70\,\% ethanol to wet the channel. Ethanol was then flushed out with DI water, followed by 30 minutes of sodium hypochlorite (NaClO) injection for disinfection. Finally, the system was rinsed and saturated with sterile DI water. The second part included injecting the bacterial culture followed by nutrient broth. The flow rate of the solutions were adjusted to provide approximately the same fluid velocity between experiments.
The experimental setup can be seen in the SI in the Figure \ref{fig:Setup}. The liquid culture of bacteria was injected into the channel from the bottom inlet for 5 hours using the flow rates shown in Table \ref{table:exp_conditions}. Right after 5 hours, the bacteria injection was stopped and the nutrient broth was injected continuously into the channel from the top inlet with the flow rates given in Table \ref{table:exp_conditions}.\newline
The mean velocities inside the microfluidic channels were calculated using the continuum equation, where the area A was calculated the channel height \textit{h} and width \textit{w}. The following equation shows the calculation for the mean velocity of nutrient broth ($v_0$) in Experiment E3 (see Table \ref{table:exp_conditions}). It is important to remember that the flow rates are calculated before any biofilm growth happens since the growth will alter the velocities. 
\begin{equation*}
    v_{0} = \frac{Q_{\text{n.broth}}}{w\cdot h} = \frac{3.5\cdot  10^{-3}\,\frac{\text{mm}^3}{\text{s}}}{0.125\,\text{mm}\cdot0.035\,\text{mm}} = 0.8 \frac{\text{mm}}{\text{s}}
\end{equation*} 
The velocity profile in the channels was calculated for a rectangular channel based on \cite{Joseph1868}. The method of obtaining velocity profiles can be found in \cite{Esfahani}. The reference shear stress, $\tau_{\text{ref}}$, was calculated along both height and width of the channels by taking into account the dynamic viscosity of the water at room temperature \cite{Perry}. $\tau_{\text{ref}}$ values at the middle points of the width and length of microfluidic channels are shown in Table \ref{table:exp_conditions}. Furthermore, the velocity profile and $\tau_{\text{ref}}$ for E3 along the height and width, and its velocity profile in 2D can be found in SI in Figures \ref{fig:vpheight}, \ref{fig:vpwidth}, \ref{fig:vp2D}, \ref{fig:shearwidth}, \ref{fig:shearheight}.  Since we calculated the shear stress before any biofilm growth is observed, we call it the reference shear stress, $\tau_{\text{ref}}$. 

\begin{table}[ht!]
    \centering
     \caption{Experimental conditions for biofilm growth in different microfluidic channels. The table includes the channel type, initial flow rates of bacterial culture (\boldmath$Q_{\text{bacteria}}$) and nutrient broth (\boldmath$Q_{\text{n.broth}}$), the mean velocity of nutrient broth  (\boldmath$v_0$), and reference shear stresses (\boldmath$\tau_{\text{ref}}$). The experiments are named and arranged in order of increasing mean velocities, except for E7 the narrow channel: with pores.}
    \resizebox{\textwidth}{!}{%
    \begin{tabular}{ccccccc}
        \toprule
        \textbf{Experiment} & \textbf{Channel Type} & 
        \makecell{\boldmath$Q_{\text{bacteria}}$ \\ (\SI{}{\micro\liter\per\second})} & 
        \makecell{\boldmath$Q_{\text{n.broth}}$ \\ (\SI{}{\micro\liter\per\second})} & 
        \makecell{\boldmath$v_0$ \\ (\SI{}{\milli\meter\per\second})} & 
        \makecell{\boldmath$\tau_{\text{ref, height}}$ \\ (\SI{}{\milli\pascal})} & 
        \makecell{\boldmath$\tau_{\text{ref, width}}$ \\ (\SI{}{\milli\pascal})} \\
        \midrule
        E1 & Narrow channel & $1.75 \cdot 10^{-4}$ & $5 \cdot 10^{-4}$ & 0.114 & 17.0 & 23.6 \\
        E2 & Narrow channel & $1.75 \cdot 10^{-4}$ & $1 \cdot 10^{-3}$ & 0.23  & 34.2 & 47.6 \\
        E3 & Narrow channel & $1.75 \cdot 10^{-4}$ & $3.5 \cdot 10^{-3}$ & 0.8   & 119.0 & 165.5 \\
        E4 & Narrow channel & $1.75 \cdot 10^{-4}$ & $7 \cdot 10^{-3}$   & 1.6   & 238   & 331 \\
        E5 & Wide channel & $3.5 \cdot 10^{-4}$  & $7 \cdot 10^{-3}$   & 0.8   & 108.6 & 150.4 \\
        E6 & Wide channel & $3.5 \cdot 10^{-4}$  & $1.4 \cdot 10^{-2}$ & 1.6   & 217.2 & 300.8 \\
        E7 & \makecell{Narrow channel \\ with pores} & $1.75 \cdot 10^{-4}$ & $3.5 \cdot 10^{-3}$ & 
        \makecell{$(channel) 0.8$ \\ $(pore\ body) 0.1$} & 
        $(pore) 12.4$ & $(pore) 17.41$ \\
        \bottomrule
    \end{tabular}%
    }
    \label{table:exp_conditions}
\end{table}

Table \ref{table:exp_conditions} shows the name of the experiment, the flow cell type used, the flow rate used during the bacterial culture injection and the nutrient broth injection, as well as the resulting nutrient broth velocities and the reference shear stress.
E1, E3, E4, E5, and E6 experiments were repeated twice to assess reproducibility. In the following sections, experiments will be indicated with an additional \_1 or \_2 added to the experiment number to denote the first or second replicate (e.g., E1\_1, E1\_2).

\subsubsection{Image Processing}\label{subsubsec4}
Image analysis was performed to estimate the fractional volume of each pore or pore channel occupied by estimated biofilm volume. Images of the microfluidic channels in the experiments were captured by a camera every 5 min and saved for processing. 
ImageJ, v1.54i was used to find the coordinates for the appropriate mask used for processing; MATLAB, was used for calculating the biofilm coverage in terms of volume in the microfluidic channels and pores. 
All images were cropped according to the mask and turned into binary files, only distinguishing between biofilm and fluid phase. 
Simple calculation of the mean value across the channel returned a value for the covering for each frame.

\subsection{Simulations}\label{subsec:methods_simulation}
\subsubsection{Biofilm model}\label{subsubsec:methods_model}
Our generic bacterial model is based on previous work and detailed in \cite{lohrmann23a}.
Here; we provide a brief summary here.
We represent bacteria as particles in a coarse-grained molecular dynamics scheme, coupled to lattice-Boltzmann hydrodynamics.
Cells are represented by $N_\text{Beads}$ = 5  spherical particles that are rigidly connected to form a rod. Here, the strength of attractive interaction between cells represents a proxy mechanism to account effectively for cohesiveness, while EPS as such is not explicitly modelled.
The translational and rotational dynamics are governed by the Langevin equation of motion.
Forces and torques arise from interactions with other cells, modeled by the Lennard-Jones potential, interactions with confining geometries, modeled by the Weeks-Chandler-Anderson potential, and interactions with an underlying fluid.
We use the lattice-Boltzmann method to solve the Navier-Stokes equations for an incompressible Newtonian fluid with bounce-back boundary conditions on no-slip surfaces and constant velocity boundary conditions at the inlet and outlet.
Particles and the fluid are coupled by friction force 
\begin{equation}
        \vb{F}_i = \bareFrict (\velFluid(\beadPos{i}) - \velBact{_i}),
        \label{eg:lb_coupling_force}
\end{equation}
where $\bareFrict$ is the friction coefficient, $\velFluid(\beadPos{i})$ the velocity of the fluid at the position $\beadPos{i}$ of particle $i$, and $\velBact{_i}$ the velocity of the particle.
The friction force is applied to the particles, and, with opposite sign, to the fluid.
Thus, the two-way coupling obeys Newton's third law of motion.
The value of the friction coefficient $\gamma$ is chosen as large as possible -- without impeding numerical stability -- to mimic the no-slip boundary condition on the surface of the bacteria.

Beyond physical interactions, biological detail is included in two ways.
Surface interactions are modeled by an algorithm for reversible bond formation.
Whenever cells are close to a surface, there is a chance that a harmonic bond is created with an ``anchor'' particle that is placed on the surface. 
Effectively, this represents biofilm adhesion.
If the force on the bond exceeds a threshold value, the bond and the anchor are removed.
Cell growth is modeled by successively increasing the distance between the particles that make up the rigid rod.
Assuming nutrient availability, growth is exponential with a constant growth rate, $r_{\text{growth}}$. 
Upon reaching a threshold cell length, a division event is triggered.
Then, the dividing cell is replaced by two daughter cells that together occupy the same volume as the mother cell did. This is called the doubling time $\tau_2$, and it is calculated as $\tau_2$ = ln(2)/$r_{\text{growth}}$ \cite{lohrmann23a}. 

In summary, the computational model is a generic model based on physical mechanisms, which it does not incorporate factors such as the nutrient availability in the experiments or the structural and rheological effects of the biofilm. The model is used for this study with the aim to qualitatively validate the governing mechanisms observed in the development of intermittent flow paths.

\subsubsection{Simulation setup}\label{subsubsec:methods_sim_setup}
\begin{figure}[H]
    \centering
    \includegraphics[width=\linewidth]{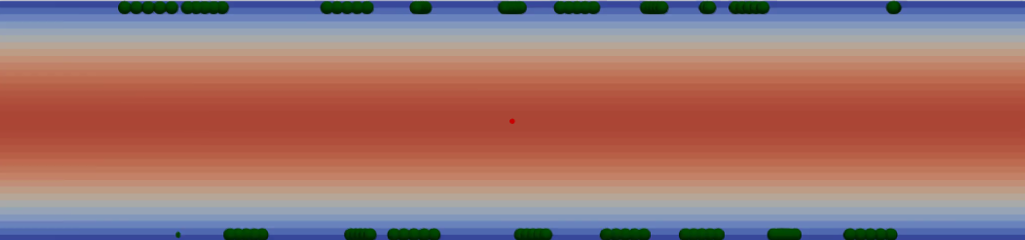}
    \caption{Two-dimensional representation of Lattice-Boltzman modeling domain consisting of a rectangular channel with the dimensions  $l_x\cross l_y  \cross l_z = \SI{150}{\micro\meter} \cross \SI{50}{\micro\meter} \cross \SI{35}{\micro\meter}$. There are no-slip boundary conditions in the \textit{y} and \textit{z} directions and along \textit{x} there is a Poiseuille flow enforced at the inlet and outlet. The flow velocity in the center of the channel is indicated by the background color. Initially, we randomly place $N_\text{bact}$ = 50 bacteria on the bottom and top surface, displayed here as dark green particles.}
    \label{fig:simsetup}
\end{figure}

Simulations were performed in a rectangular channel of dimensions $l_x\cross l_y  \cross l_z = \SI{150}{\micro\meter} \cross \SI{50}{\micro\meter} \cross \SI{35}{\micro\meter}$.
This geometry for the Lattice-Boltzman simulations had to be  significantly smaller than the experimental setup to make simulations computationally feasible, but still allows qualitative comparison.
We place walls with a no-slip boundary condition in the $y$ and $z$ direction.
Along $x$, at the inlet and outlet, we enforce the velocity profile for Poiseuille flow in a rectangular channel with a mean velocity of $\expval{\velFluid} = \SI{1}{\milli\meter\per\second} \vb{e}_x$, where $\expval{\cdot}$ denotes an average over the inlet surface and $\vb{e}_x$ the unit vector along $x$.
Initially, we randomly place $N_\text{bact}$ = 50 bacteria on the top and bottom surfaces to initiate the biofilm growth as shown in Figure \ref{fig:simsetup} in SI. We performed the simulations with varied biofilm cohesiveness while maintaining a constant growth rate and inflow velocity. Varying the inflow velocity is not feasible in our simulation setup, as time-scale separation between bacterial growth (slow) and movement of bacteria in the flow (fast) must be maintained and therefore the inflow velocity cannot be reduced significantly.
Particles that exceed the domain boundaries along $x$ are removed from the simulation.

\section{Results and discussion}\label{sec:results}
\begin{figure}[h]
    \centering
    \begin{subfigure}{0.49\textwidth}
        \centering
        \includegraphics[width=\textwidth]{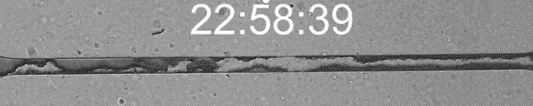} 
        \caption{}
        \label{fig:a}
    \end{subfigure}
    \hfill
    \begin{subfigure}{0.49\textwidth}
        \centering
        \includegraphics[width=\textwidth]{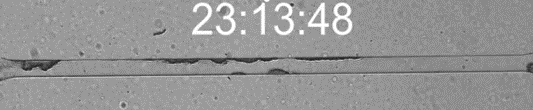} 
        \caption{}
        \label{fig:b}
    \end{subfigure}
    \caption{Two events from experiment E3 ($v_0=0.8$mm/s, w = 0.125 mm): (a) showing biofilm accumulated in the flow channel, and (b) the flow channel mostly free of biofilm approximately 15 min later after a major detachment event. The numbers in the figure represent timestamps in hours:minutes:seconds.}
    \label{fig:biofilmcoloration}
\end{figure}

We observed biofilm growth in microfluidic channels of different sizes and designs under varying flow rates. Our findings indicate that constant flow can lead to preferential flow paths. This finding which is consistent with previous studies \cite{Zhou2020, Sharp2025, Kurz2022, Bottero2013, Stewart2001}. In particular, we observed that the formation of preferential flow paths is primarily influenced by the mean velocity, while it remains largely unaffected by the physical dimensions of the microfluidic channel.
In all experiments, as shown in the videos in the SI, the biofilm became visible as a dark discoloration approximately 4–8 hours after the injection of nutrient broth. The experiments demonstrated that biofilm accumulation and detachment alternated, with biofilm accumulation through growth (and possibly re-attachment) occurring fairly slowly, and biofilm detachment occurring faster, almost suddenly, likely through biofilm sloughing events. These observations agree with those of \citet{Kurz2022}. For instance, as indicated in Figure \ref{fig:a} the accumulation of biofilm led to the narrowing of the microfluidic channel; Figure \ref{fig:b} demonstrates that the flow path reopened shortly thereafter due to biofilm detachment.

In section~\ref{ss::qualitative}, we discuss the experimental results from a qualitative point of view and in  section~\ref{ss::quantitative}, we quantitatively analyse the microfluidic experiments using image processing to conclude on flow characteristic impacts on the intermittency of flow paths during biofilm growth. Furthermore the results of numerical simulations are compared to the experimental findings in section~\ref{ss::simresults}.  

\subsection{Qualitative analysis of the experiments}\label{ss::qualitative}
In this section, we discuss results qualitatively by separating the effects of (i)~the channel width,  (ii)~the average velocity in the channel, and (iii)~the addition of pore bodies on the characteristics of biofilm growth.   

\subsubsection{Effect of channel width}
To investigate the effect of the channel width on the characteristics of biofilm growth, we compared experiments where the mean initial velocity is the same, but the channel width, $w$, is different. 
More precisely, we compare experiment E3 ($w = 0.125$~mm) to E5 ($w = 0.25$~mm), both with an average velocity of 0.8~mm/s in the channel as well as experiment E4 ($w = 0.125$~mm) to E6 ($w = 0.25$~mm), both with an average velocity of 1.6~mm/s (cf. Table~\ref{table:exp_conditions}).

\begin{figure}[H]
    \centering
    \begin{minipage}[b]{0.45\textwidth}
        \centering
        \includegraphics[width=\textwidth]{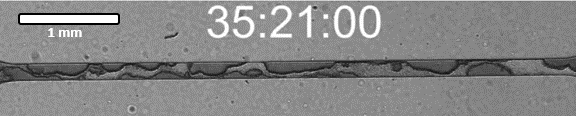}
        \subcaption{}  % Subcaption for a
    \end{minipage}
    \hfill
    \begin{minipage}[b]{0.45\textwidth}
        \centering
        \includegraphics[width=\textwidth]{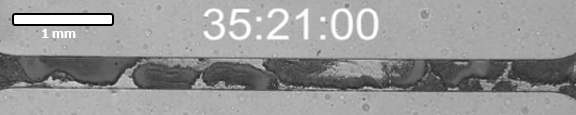}
        \subcaption{}  % Subcaption for b
    \end{minipage}
    
    \vskip\baselineskip  % Adds space between the rows
    
    \begin{minipage}[b]{0.45\textwidth}
        \centering
        \includegraphics[width=\textwidth]{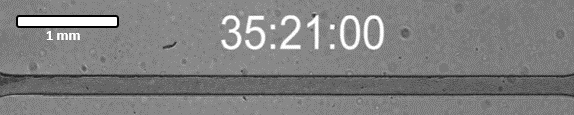}
        \subcaption{}  % Subcaption for c
    \end{minipage}
    \hfill
    \begin{minipage}[b]{0.45\textwidth}
        \centering
        \includegraphics[width=\textwidth]{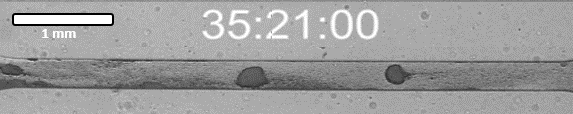}
        \subcaption{}  % Subcaption for d
    \end{minipage}

    \caption{The comparison of channel widths (w) under constant mean velocities: a) E3 ($v_0=0.8$mm/s, w = 0.125 mm), b) E5 ($v_0=0.8$mm/s, w = 0.25 mm), c) E4 ($v_0=1.6$mm/s, w = 0.125 mm), d) E6 ($v_0=1.6$mm/s, w = 0.25 mm).}
    \label{fig:E3E5E4E6}
\end{figure}

Despite the increase in width, the behavior of the biofilm is similar in these experiments (Figure \ref{fig:E3E5E4E6}). In the experiments E3 and E5 experiments ($v_0$ = 0.8 mm/s), we observed biofilm growth with preferential flow paths, whereas in the experiments E4 and E6 experiments ($v_0$ = 1.6 mm/s), the intermittent regime was only for a short duration. Later, the ruptures in the biofilm began to appear and we did not observe a net biofilm accumulation afterward. All these observations can be seen in the SI videos \textit{V-E3.1.mp4, V-E3.2.mp4, V-E4.1.mp4, V-E4.2.mp4, V-E5.mp4} and \textit{V-E6.1.mp4, V-E6.2.mp4}. We divided these four experiments into two groups based on the mean velocities, suggesting that channel width does not significantly influence biofilm growth; rather, the biofilm behavior changes with variations in mean velocity. The impact of variations in velocity will be addressed in more detail in the following section. 
\subsubsection{Effect of velocities}
To investigate the effect of the mean initial velocities on the characteristics of biofilm growth, we compare experiments where the geometry of the microfluidic channel is the same, but the applied flow rate of the nutrient broth is altered, resulting in varying mean initial velocities, $v_{0}$.
We analyze two sets of experiments: Set~1 includes the experiments E1 ($v_0=0.114$mm/s), E2 ($v_0=0.23$mm/s), E3 ($v_0=0.8$mm/s), E4 ($v_0=1.6$mm/s), all with a channel width of $w=0.125$mm. 
Set~2 includes the experiments E5 ($v_0=0.8$mm/s) and E6 ($v_0=1.6$mm/s), both with a channel width of $w=0.25$mm.
The corresponding experimental parameters are given in Table~\ref{table:exp_conditions}. 

We performed two replicates of experiment E1 ($v_0=0.114$mm/s). In the E1.1 and E1.2 experiments, we did not observe any biofilm development. These experiments were conducted at the lowest flow rate, resulting in a supposedly limiting nutrient supply (0.24 µg/min of nutrient broth powder), see also SI Table \ref{tab:nutrients}. To confirm this hypothesis, we performed another experiment (E1.3) with the same mean velocity ($v_0=0.114$mm/s) by increasing the nutrient broth concentration to 0.48 µg/min, i.e., the same supply as it was used in the E2 experiment. These experiments can be seen in the SI videos \textit{V-E1.1.mp4}, \textit{V-E1.2.mp4}, and \textit{V-E1.3.mp4}. The experiment E1.3 showed a biofilm development which did not last for long. There are multiple mechanisms interacting and to develop an approach to explain this, firstly we need to reconsider the effect of shear. In the E1 experiments, the low flow rate ($v_0=0.114$mm/s) resulted in low shear stress ($\tau_{\text{ref, height}}$ = 17 mPa, $\tau_{\text{ref, width}}$ = 23.6). Therefore,the shear stress was not sufficiently high to result in the development of distinct flow paths. On the other hand, we suppose that the cohesiveness of the biofilm grown under low-shear-stress conditions is less than under higher shear. This was also reported, although for another species, by \citet{Stoodley2002} and \citet{Fanesi2021}, who also observed that biofilms under lower shear-stress tend to produce less biomass. The relationship between the biofilm's cohesiveness and shear stress will be discussed in more detail in Section 3.2.2. Additionally, under low shear conditions, it is known that biofilms have been described to adapt their internal structures to these conditions in order to facilitate nutrient transfer to deeper layers. This adaption increases their porosity and permeability, resulting in more fluffy biofilm structure, with decreased density \cite{Beyenal2002, Stoodley1998, Moreira2013, PaulE2012, Characklis1981rep}.
We assume that as a result, in the E1 experiments shear stress was small enough and nutrient supply was low enough to not force flowpath development. In the absence of flow paths, we assume that bacterial cells are filtered by the biofilm, preventing sustainable biofilm growth further downstream.
 In summary, we note that the E1 experiments fail to provide the conditions for observing intermittent regime.
 
\begin{figure}[ht!]
    \centering
    \begin{minipage}[b]{0.45\textwidth}
        \centering
        \includegraphics[width=\textwidth]{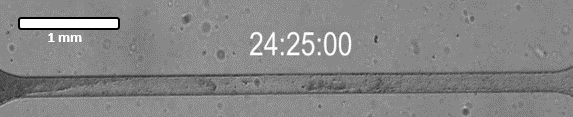}
        \subcaption{}
    \end{minipage}
    \hfill
    \begin{minipage}[b]{0.45\textwidth}
        \centering
        \includegraphics[width=\textwidth]{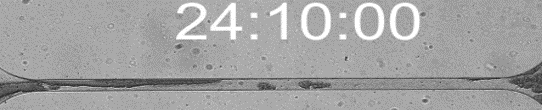}
        \subcaption{}
    \end{minipage}
    
    \caption{The comparison of velocities in the narrow flow channel width. a) E1 ($v_0=0.114$mm/s, w = 0.125 mm), b) E2 ($v_0=0.23$mm/s, w = 0.125 mm).}
    \label{fig:E1vsE2}
\end{figure}

In E2 ($v_0=0.23$mm/s in narrow channel) (see the video \textit{V-E2.mp4}), E3 ($v_0=0.8$mm/s in narrow channel), and E5 ($v_0=0.8$mm/s in wide channel), we can observe the longer-term interaction of biofilm growth and shear during intermittent regime as shown in Figure \ref{fig:E1vsE2}b. The established regime in these experiments (E2, E3, and E5) meet conditions which, as stated by \citet{Stoodley1998}, indicate a denser and more uniform structure of biofilm development to adapt to increasing shear, rather than the looser or more porous biofilm in the E1 experiment. In these experiments, the shear forces were strong enough to open new flow paths, while the biofilm, supported by sufficient nutrient availability, has grown and developed resistance to these forces.
In the E4 ($v_0=1.6$mm/s in narrow channel) and E6 ($v_0=1.6$mm/s in wide channel) experiments (see Fig \ref{fig:E3E5E4E6}), biofilm development could not sustain as we indicated in the previous section.
These findings show that biofilm and shear stress interact in a dynamic equilibrium in experiments E2, E3, and E5, while shear stress dominates over biofilm growth and disrupted the dynamic equilibrium in experiments E4 and E6. In these two experiments, there is a brief period of growth, which is quickly followed by a shearing-off period once the shear stresses are sufficiently strong. As observed in the V-E4.2 video, the biofilm completely detached at the 35th hour and did not regrow, despite the experiment continuing until the 48th hour.

Based on these observations, we can divide the experiments into three groups. The first group includes the E1 experiments ($v_0 = 0.114 \, \text{mm/s}$ in narrow channel), where biofilm growth was limited rather by nutrients than by shear stress. The second group comprises the E2 ($v_0 = 0.23 \, \text{mm/s}$ in narrow channel), E3 ($v_0 = 0.8  \, \text{mm/s}$ in narrow channel), and E5 experiments ($v_0 = 0.8  \, \text{mm/s}$ in wide channel), which showed an intermittent growth/flow regime with a dynamic balance between biofilm growth and shear forces. The third group includes the E4 ($v_0 = 1.6  \, \text{mm/s}$ in narrow channel) and E6 experiments ($v_0 = 1.6  \, \text{mm/s}$ in wide channel), where this balance was lost due to high shear. We have related the formation or disruption of this dynamic balance to the term critical shear stress, which represents the equilibrium of attachment and detachment, as defined by \citet{Stoodley2001} and \citet{Nejadnik2008}. Critical shear stress is defined as an increase in detachment relative to attachment. That is why we can suggest that attachment and detachment are still in a dynamic balance in the second group of experiments (E2, E3, and E5) based on this definition. Biofilm growth with the flow paths was also observed for a period in the third group of experiments. However, at some point, the dynamic balance between attachment and detachment was disrupted due to the dominance of shear-induced detachment, which the biofilm could not withstand, ultimately preventing any further net biofilm accumulation.
In addition to the interaction between shear stress and biofilm growth, we observed in the E2, E3, and E5 experiments that the flow paths did not always form in the same position and shape, but rather showed changes and shifts in their positions during the experiment as in the models of \citet{Bottero2013}, where decay and lysis are considered. Therefore, we can also assume that biofilm decay, lysis, and matrix degradation were present in our experiments and can be responsible for the biofilm detachments.

\subsubsection{Effect of the pore geometry}
\begin{figure}[H]
    \centering
    \begin{subfigure}{0.49\textwidth}
        \centering
        \includegraphics[width=\textwidth]{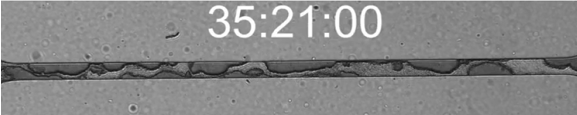} 
        \caption{}
        \label{fig:A2cellgrowth}
    \end{subfigure}
    \hfill
    \begin{subfigure}{0.49\textwidth}
        \centering
        \includegraphics[width=\textwidth]{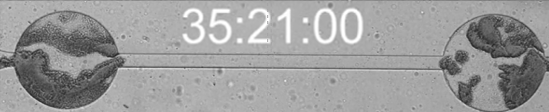} 
        \caption{}
        \label{fig:b1cellgrowth}
    \end{subfigure}
    
    \begin{subfigure}{0.5\textwidth}
        \centering
        \includegraphics[width=\textwidth]{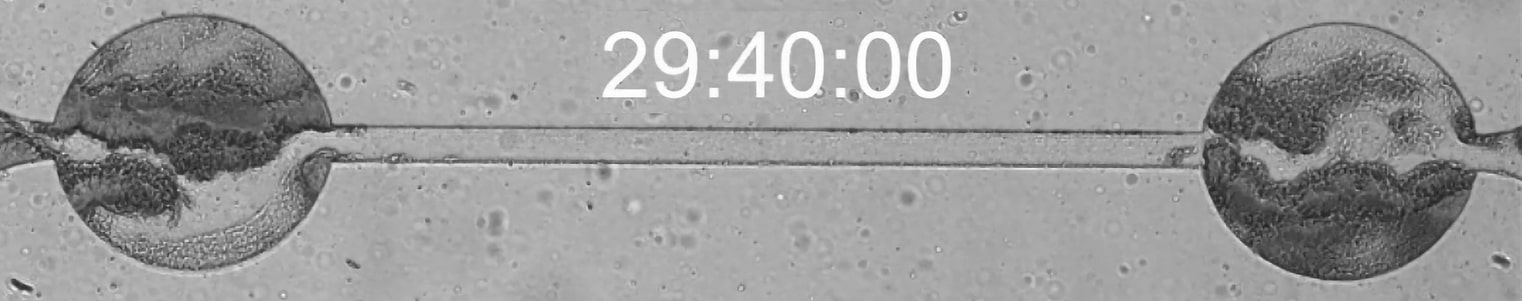}
        \caption{}
        \label{fig:diflimit}
    \end{subfigure}

    \caption{Comparison of E3 ($v_0=0.8$mm/s, w = 0.125 mm) and E7 experiments ($v_0=0.8$mm/s, w = 0.125 mm) with pore bodies. (c) Due to diffusion limitation, more biofilm accumulation, appearing as a darker discoloration, can be observed toward the center of the pore body, while less biofilm accumulates near the pore body boundaries.}
    \label{fig:E1vsE7}
\end{figure}

To investigate the effect of the pore geometry on the characteristics of biofilm behavior, we used flow channels, where the pore bodies were added to the microfluidic channels, i.e., resembling a setting of pore bodies and pore throats. In this regard, we compared the experiment E7 (narrow channel with pores) with the experiment E3 (narrow channel) both with a mean velocity of 0.8~mm/s in the channel as shown in Figure~\ref{fig:E1vsE7}. In experiment E7, we observed that growth mainly occurred in the pore bodies, with occasional growth in the channel, even if this was very short-lived. This observation can be seen in the video \textit{V-E7.mp4} in the SI. This observation aligns with the observation in \cite{Kim2000} that biofilms are more commonly seen to grow in the pore bodies.  Based on these, we concluded that the tendency of biofilm to grow in the pore bodies resulted from the lower shear stresses ($\tau_{\text{ref, height}}$ = 119 mPa in the channel, $\tau_{\text{ref, height}}$ = 12.4 mPa in the pore bodies) in the pore bodies compared to the channel. This comes as a result of more challenging growth conditions in terms of shear stresses  in the channel for biofilm. This is due to the fact that the fluid slows down when reaching the pore body, thus reducing the shear stress. The biofilm has adapted to the lower shear stress in the bodies and can no longer withstand the higher shear stress in the channel.
Our result here supports the results of \citet{Zhou2020} in terms of biofilm's tendency to grow where shear forces are lower as well as the results of \citet{Fanesi2021}, which can be interpreted in terms of lower cohesiveness of the biofilm grown in the pore body. \\
In addition to previous observations, the biofilm often grows more toward the center of pore bodies than across the entire area of the pore body, as seen in the Figure~\ref{fig:diflimit}.
This situation arises from the diffusion limitation at the upper and lower parts of the pore bodies, and with that higher nutrient availability near the middle regions.
In summary, we can conclude that the biofilm behavior is significantly altered by adding bodies such that under otherwise identical flow and transport conditions the channels/throats are almost fully cleared of biofilm growth.

\subsection{(Semi-) Quantitative Analysis}\label{ss::quantitative}
\subsubsection{Image Processing results}
\label{sec:results_image_processing}
\begin{figure}[h]
    \centering
    \includegraphics[scale=0.22]{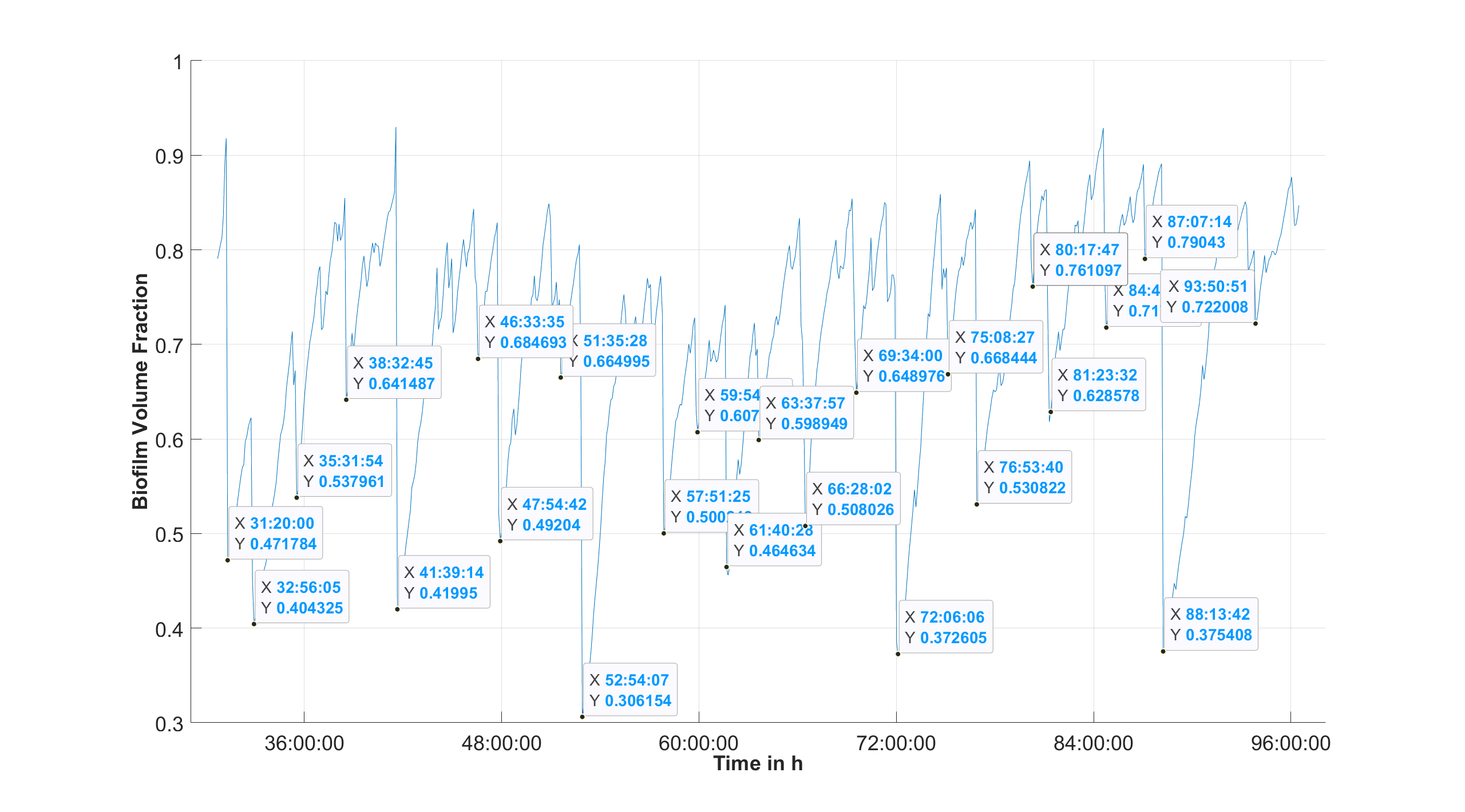}
    \caption{Temporal evolution of the coverage ratio of the biofilm in the first trial of E3.1 ($v_0 = 0.8 \, \text{mm/s}$ in narrow channel) for calculation of the frequency of detachment events. The selected points are shown, and these points are where large biofilm fragments begin to accumulate immediately after detachment.}
    \label{fig:im_pro_markers}
\end{figure}

Image processing was used to further analyze the experimental observations. This was aimed at enabling a quantitative analysis and comparison of the frequency of detachment events. While the biofilm accumulates dynamically, the frequency of the resulting flow paths is also equal to the frequency of detachment events. The  biofilm coverage, which serves as a good proxy for estimating the biofilm growth rate, was also obtained through image processing. As an example, Figure \ref{fig:im_pro_markers} shows the temporal evolution of the biofilm coverage in the first trial of Experiment~3 (E3.1). This plot in the main part of this manuscript only include the part of the experiment that contain the intermittent regime. The plots for the full length of each experiment are included in the SI.

To estimate the frequency of detachment events, a set of images from each experiment was selected, including a section where the preferential flow paths are clearly observable (for example, images between hours 12 and 90 in the E3 experiment video, see V-E3.1.mp4), and manually analyzed. For this analysis, each major detachment point was selected on each plot. Major points correspond to the troughs of peaks where changes in biofilm coverage are more than approximately 10\,\%. For instance, the point at 32:56:05 with 40\,\% biofilm coverage increment was selected, representing the beginning of a peak that rises to approximately 70\,\% biofilm coverage as shown in Figure \ref{fig:im_pro_markers}. Points with less than 10\,\% coverage increments were considered noise or minor detachments.  Figure~\ref{fig:im_pro_markers} illustrates the identified detachment points within the selected part of E3. As a result, the frequency of detachment events was calculated by first determining the time intervals between manually selected detachment points (e.g., the interval between 32:56:05 and 35:31:54 corresponds to 9,324 seconds). The reciprocal of the harmonic mean of all intervals gives the frequency of detachment events.

The net biofilm growth rate (coverage/s), which indicates how fast the biofilm covers the channel on average, was obtained from the slope of each main peak in these graphs (peaks with a loss in biofilm coverage of more than 0.10). 
This method was repeated across different experiments to assess variations in the frequency of detachment events. A summary of the results is provided in Table~\ref{tab:im_pro_freq}, while all related graphs are available in the Supporting Information (SI).

\begin{table}[h]
    \centering
        \caption{Results for frequencies of detachment events and net biofilm growth rates in different experiments. Coverage refers to the estimated fraction of the total surface area that is occupied by the biofilm in terms of volume. For experiments without any values, calculations could not be made, as no net biofilm accumulation was observed.}
    \resizebox{\textwidth}{!}{%
    \begin{tabular}{ccccc}
        \toprule
        \textbf{Experiment} & \textbf{Channel Type} & \textbf{\makecell{Velocity \\ (\SI{}{\milli\meter\per\second})}} & 
        \textbf{\makecell{Frequency of \\ detachment events \\ (\SI{}{\per\second})}} & 
        \textbf{\makecell{Net biofilm growth rate \\ (coverage/\SI{}{\second})}} \\
        \midrule
        E1.1    & Narrow channel  & 0.114 & - & - \\
        E1.2    & Narrow channel  & 0.114 & - & - \\
        E1.3    & Narrow channel  & 0.114 & - & - \\
        E2    & Narrow channel  & 0.23  & $1.23 \cdot 10^{-4} \pm 7.5 \cdot 10^{-5}$ &  $3.32 \cdot 10^{-5} \pm 1.3 \cdot 10^{-5}$  \\
        E3.1  & Narrow channel  & 0.8   & $1.52 \cdot 10^{-4} \pm 1.1 \cdot 10^{-4}$ & $4.0 \cdot 10^{-5} \pm 8.3 \cdot 10^{-6}$ \\
        E3.2  & Narrow channel  & 0.8   & $1.45 \cdot 10^{-4} \pm 6.6 \cdot 10^{-5}$ & $4.10 \cdot 10^{-5} \pm 8.0 \cdot 10^{-6}$ \\
        E4.1  & Narrow channel  & 1.6   & $2.25 \cdot 10^{-4} \pm 1.5 \cdot 10^{-4}$ & $4.71 \cdot 10^{-5} \pm 2.12 \cdot 10^{-5}$ \\
        E4.2  & Narrow channel  & 1.6   & $2.32 \cdot 10^{-4} \pm 1.7 \cdot 10^{-4}$ & $5.40 \cdot 10^{-5} \pm 1.9 \cdot 10^{-5}$ \\
        E5  & Wide channel & 0.8   & $1.37 \cdot 10^{-4} \pm 9.8 \cdot 10^{-5}$ & $2.65 \cdot 10^{-5} \pm 5.70 \cdot 10^{-6}$ \\
        E6.1  & Wide channel & 1.6   & - & - \\
        E6.2  & Wide channel & 1.6   & - & - \\
        E7    & \makecell{Narrow channel \\ with pores} & 0.8   & - & - \\
        \bottomrule
    \end{tabular}%
    }
    \label{tab:im_pro_freq}
\end{table}

In the previous section, we divided the experiments into three groups based on how the biofilm growth developed. For the first group, containing different replicates of E1 ($v_0 = 0.114 \, \text{mm/s}$ in narrow channel), we were unable to do image processing since there were no apparent flow paths. Therefore, Table~\ref{tab:im_pro_freq} does not include a frequency calculation for this experiment. However, it is rather obvious how the frequency of detachment events of the second and third-group experiments differ from one another. The mean velocity of nutrient broth solution in the E2 experiment ($v_0 = 0.23 \, \text{mm/s}$ in narrow channel) is less than in E3 ($v_0 = 0.8 \, \text{mm/s}$ in narrow channel) and E5 ($v_0 = 0.8 \, \text{mm/s}$ in wide channel). We found that the frequency of detachment events in E2 was also lower than the frequency of detachment events in E3 and E5, which implies that the resulting flow paths were also less frequent. In the E2 experiment, the observed lower frequency of flow paths and slower biofilm growth rate are due to the smaller amount of available nutrients. Due to nutrient limitation, the biofilm grew while developing a weaker structure. However, the reason why the frequency is lower in E2, but still close to E3, is that the biofilm is still growing in a dynamic equilibrium state. In the third group of experiments, containing E4 and E6, the mean velocity of nutrients was higher ($v_0 = 1.6 \, \text{mm/s}$) than that in the second group ($v_0 = 0.23$ or $0.8 \, \text{mm/s}$). Therefore, this caused the dynamic equilibrium to break down after some time. This is because, we found a higher frequency of detachment events and flow paths with the similar observations of the results from \citet{Kurz2022} and higher coverage rates in these experiments, namely, biofilm accumulated faster but at the same time the detachment events were observed more frequently. Consequently, the balance between shear stress and biofilm growth was disrupted after some time, and neither flow paths nor biofilm growth were observed anymore. That's why, the frequency of detachment events and flow paths in the E4 and E5 was calculated only until the total biofilm detachment. 

\subsubsection{Dimensional analysis}
We did a dimensional analysis for our experiments to better understand the biofilm growth and to classify the experiments. In this analysis, we took into account the channel height as the characteristic length, the water density, biofilm cohesiveness, water dynamic viscosity, and the nutrient broth velocity as dimensional parameters. We arrived at the Reynolds number (Re) and the Euler-Cohesive number (Eu\textsubscript{co}) as the two dimensionless numbers according to Buckingham's Pi theorem  \cite{Buckingham1914}.

\begin{figure}[H]
    \centering
    \includegraphics[scale=0.5]{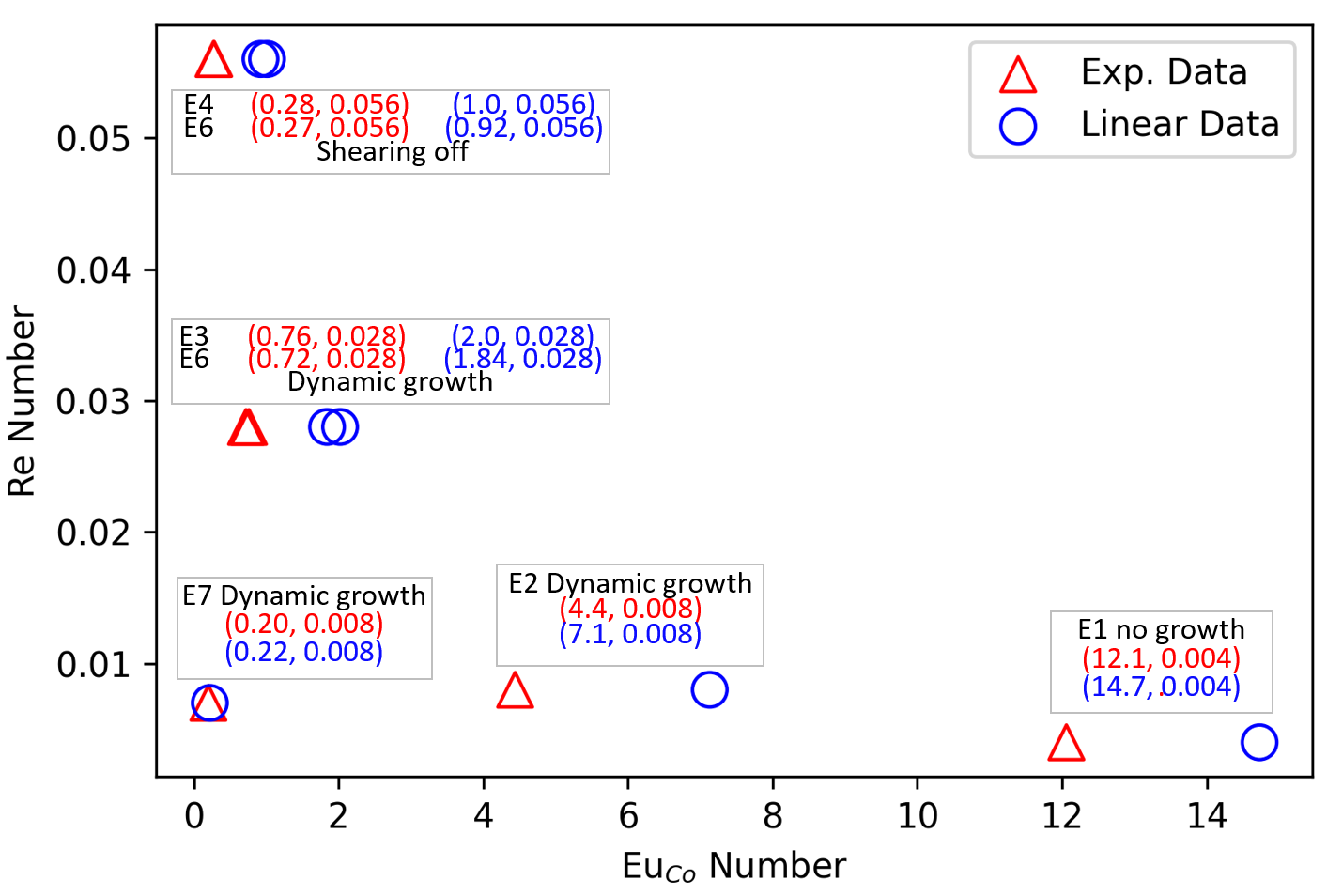}
    \caption{Re and Eu\textsubscript{co} numbers for the experiment with narrow and wide flow channels, including E1 ($v_0 = 0.114 \, \text{mm/s}$ in narrow channel), E2 ($v_0 = 0.23 \, \text{mm/s}$ in narrow channel), E3 ($v_0 = 0.8 \, \text{mm/s}$ in narrow channel), E4 ($v_0 = 1.6 \, \text{mm/s}$ in narrow channel), E5 ($v_0 = 0.8 \, \text{mm/s}$ in wide channel), and E6 ($v_0 = 1.6 \, \text{mm/s}$ in wide channel). The values in parentheses are the x and y value. For example, (x, y) = (9.4, 0.056). The x-values of each experiment include both the Eu\textsubscript{Co, linear} calculated using the linear Co formula and the Eu\textsubscript{Co, log} calculated using the exponential Co formula.
}
    \label{fig:dimensional}
\end{figure}

A sample calculation for the Re numbers is shown below.
The calculated Reynolds numbers in the experiments were always below~1 as seen in Table~\ref{table:calc_values}, meaning that a creeping flow was present at the beginning of the injections.
\[
\mathrm{Re} = \frac{\mathrm{v_{nb}} \cdot \mathrm{h} \cdot \rho}{\mu} = \frac{0.8\,\mathrm{mm/s} \cdot 35\,\mu\mathrm{m} \cdot 1000\,\mathrm{kg/m^3}}{10^{-3}\,\mathrm{kg/(m \cdot s)}} = 2.8 \cdot 10^{-2}
\]
The Eu\textsubscript{co} number can be compared to the dimensionless Eu number. The only difference between Eu\textsubscript{co} and Eu is that the cohesive resistance of the biofilm is used instead of the pressure difference in Eu. The units of both terms are the same (Pa). The Eu\textsubscript{co} number represents the ratio of biofilm cohesiveness to the inertial forces of the fluid, assisting us in comprehending how the cohesive forces in the biofilm interact with the flow dynamics. \\
Biofilm is soft and gelatinous, in particular, \citet{MartinRoca2023} have stated that P. fluorescens biofilms exhibit soft solid characteristics and have a viscoelastic structure. For these reasons, cohesiveness is difficult to determine, so parameters like Young's modulus and apparent shear modulus (\( E_{\text{app}} \)) have been studied to estimate its cohesive character \cite{Ahimou2007, MartinRoca2023}. Stoodley et al. established for Pseudomonas aeruginosa a correlation between the shear stress that induces biofilm detachment and \( E_{\text{app}} \), which serves as a measure of biofilm rigidity, by exposing biofilms to varying shear forces \cite{Stoodley2002}. The EPS composition of P. aeruginosa and P. fluorescens species is similar, and therefore, they are expected to exhibit similar mechanical behavior \cite{MartinRoca2023}. They obtained two relationships between \( E_{\text{app}} \) and the shear stress, $\tau$,  applied to the biofilm - one linear and one exponential — as given in the equations, respectively:
\begin{equation}
    E_{\text{app}} = 10.8 \tau + 7.8
\end{equation}
\begin{equation}
    E_{\text{app}} = 30.3 \tau^{0.58}
\end{equation}
    Shear stress is given here in Pa, the same holds for \( E_{\text{app}} \). 
  In this study, we used these empirical formulas in our calculations to estimate cohesiveness, Co, also in Pa instead of \( E_{\text{app}} \). In these calculations, we used $\tau_{\text{ref, height}}$ as the applied shear stress, since channel height is the relevant dimension for the shear stress. After estimating the cohesiveness, i.e. \( E_{\text{app}} \)  for all the experiments, in the end, we calculated the Eu\textsubscript{co} as below. Numbers refer examplary to experiment E3, while using Co as calculated by the linear formula.
\[
\mathrm{Eu}_{\text{co}} = \frac{\mathrm{Co}}{\rho \cdot (\mathrm{v}_{\text{nb}})^2} = \frac{1.29\, \si{\pascal}}{1000\, \si{\kilogram\per\cubic\meter} \cdot (0.8\, \si{\milli\meter\per\second})^2} = 2.02
\]
\\
All experiments' Re number,  Co values calculated by the linear or the exponential formulas, and Eu\textsubscript{co} are presented in Table~\ref{table:calc_values}. As observed in previous studies \cite{Stoodley2002, Fanesi2021}, we also found that biofilms grown under higher shear stress show a more cohesive structure. For example, in a section of the E6 experiment (see in \textit{V\textsubscript{Rolling}}), we observed that despite detaching from the surface, the biofilm exhibited a rolling-like movement due to its high cohesive resistance, gradually shrinking and eventually exiting the microfluidic system. This by the way explains part of the noise observed in Figure~\ref{fig:im_pro_markers}.
\\
Figure~\ref{fig:dimensional} provides an overview of how the different experiments and their characteristic behaviour correlate with the Re and Eu\textsubscript{co} numbers. Based on the data, it was observed that high Re numbers and low Eu\textsubscript{co} numbers negatively affect the net biofilm accumulation, leading to detachment of the biofilm. In this case, the biofilm cannot withstand the shear forces. On the other hand, very low Re numbers may result in high Eu\textsubscript{co} numbers, indicating resistance of the biofilm to flow-induced shear. However, in this scenario, the necessary nutrients cannot be supplied for biofilm accumulation. Obviously, biofilms that grow with an intermittent regime are observed at the intermediate values, giving proof that there exists a transition region where intermittency is characteristic. Whether the Eu\textsubscript{co} numbers are estimated based on the linear approximation of cohesiveness or on the exponential is not really important for the observations in our experiments and corresponding conclusions above.

\begin{table}[ht!]
\centering
\caption{Calculated values of flow parameters, including the the initial mean velocity of nutrient broth ($v_0$), the cohesiveness calculated from the linear empirical formula \(Co_{\text{linear}} \) and the exponential empirical formula \(Co_{\text{log}} \), initial Reynolds number (Re), and the Euler-Cohesion number ((Eu\textsubscript{co, linear})) calculated with \(Co_{\text{linear}} \) and ((Eu\textsubscript{co, log})) calculated with (Eu\textsubscript{co, log}) for different experiments.}
\resizebox{\textwidth}{!}{%
\begin{tabular}{cccccccc}
\toprule
\textbf{Experiment} & \textbf{Channel Type} & 
\textbf{\makecell{$v_0$ \\ (\SI{}{\milli\meter\per\second})}} & \textbf{\makecell{\(Co_{\text{linear}} \) \\ \si{\pascal}}} & \textbf{\makecell{\( Co_{\text{log}} \) \\ \si{\pascal}}} & 
\textbf{inital Re} & 
\textbf{\makecell{Eu\textsubscript{co, linear}}} & \textbf{\makecell{Eu\textsubscript{co, log}}} \\
\midrule
E1 & Narrow channel & 0.114 & 0.19 & 0.16 & $4 \cdot 10^{-3}$ & 14.73 & 12.06 \\
E2 & Narrow channel & 0.23 & 0.38 & 0.24  & $8 \cdot 10^{-3}$ & 7.13 & 4.44 \\
E3 & Narrow channel & 0.8 & 1.29 & 0.48  & $2.8 \cdot 10^{-2}$ & 2.02 & 0.76 \\
E4 & Narrow channel & 1.6 & 2.58 & 0.72  & $5.6 \cdot 10^{-2}$ & 1.01 & 0.28 \\
E5 & Wide channel & 0.8  & 1.18 & 0.46 & $2.8 \cdot 10^{-2}$ & 1.84 & 0.72 \\
E6 & Wide channel & 1.6  & 2.35 & 0.69 & $5.6 \cdot 10^{-2}$ & 0.92 & 0.27 \\
E7 & \makecell{Narrow channel \\ with pores} & 0.8  & 0.14 & 0.13 & \makecell{(pore body) $7 \cdot 10^{-3}$ \\ (channel) $2.8 \cdot 10^{-2}$} & 0.22 & 0.20 \\
\bottomrule
\end{tabular}
}
\label{table:calc_values}
\end{table}

\subsection{Simulation results}\label{ss::simresults}
\begin{figure}
    \centering
    \includegraphics[width=0.8\linewidth]{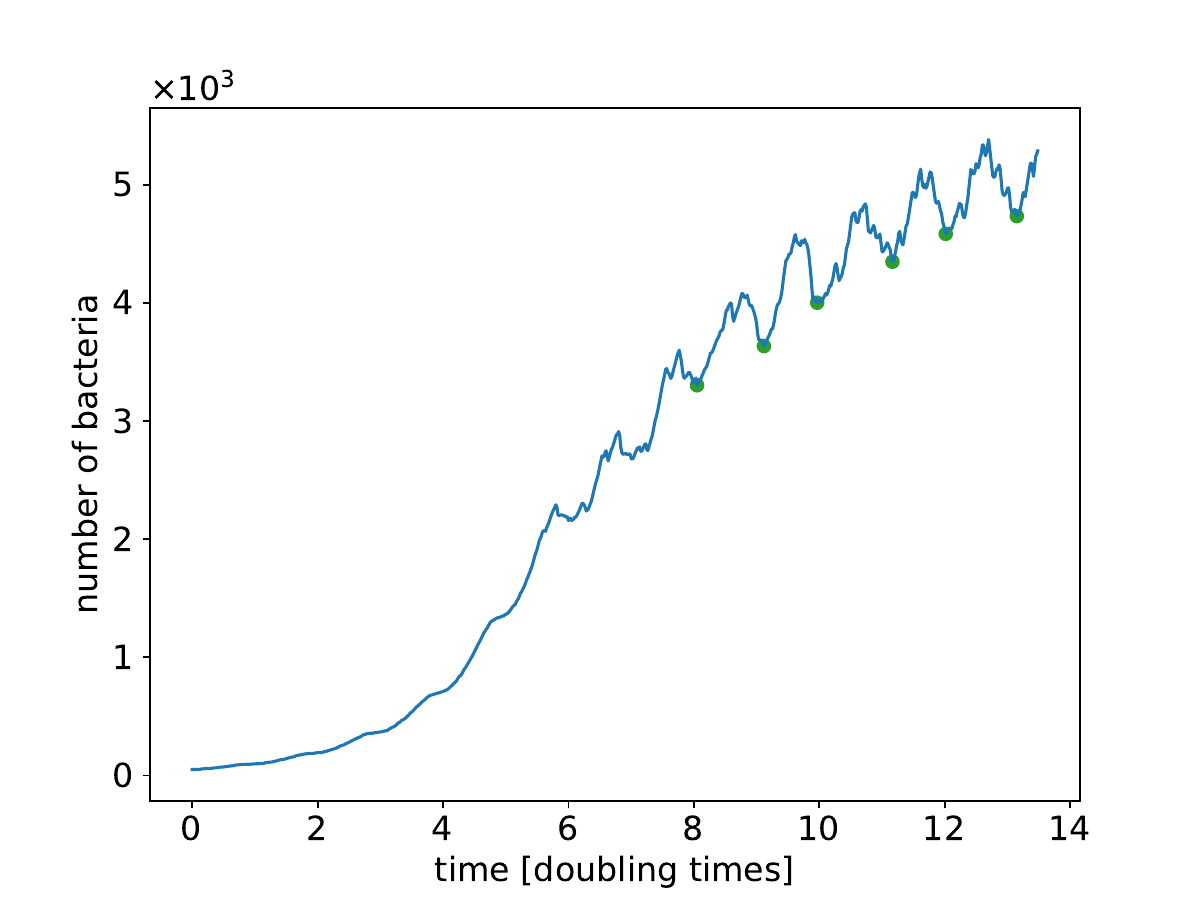}
    \caption{Number of bacteria in the simulation box as a function of time for one exemplary simulation at $\text{Co} = \SI{3}{\pascal}$. Green circles show the observed detachment events. Even before a quasi steady state is established, major detachment events occur at regular intervals.}
    \label{fig:sim_nbact_vs_time}
\end{figure}

From the results in Section~\ref{sec:results_image_processing} we hypothesize that the force balance between shear forces and biofilm cohesion only plays a minor role in determining the frequency of detachment events, whereas nutrition is the main driver.
In simulations, we can decouple the two factors to test the hypothesis.
Therefore, we performed simulations at constant growth rate and inflow velocity, while varying the biofilm cohesion, i.e., the strength of attractive interaction between cells.
A typical result for number of bacteria as a function of time is shown in Fig.~\ref{fig:sim_nbact_vs_time} and an animated video can be seen in the \textit{V-Simulation.mp4} in the SI. The growth of the cells, how doubling occurs, as well as the detachment of the cells can be seen in detail. 

Comparing this figure to the experimental results in Fig. \ref{fig:im_pro_markers}, we notice that the detachment events, i.e., drops in the number of bacteria, are much less pronounced and less sharp in the simulation.
This is because in the simulations, a large population of bacteria remains in the corners of the channel and only in the middle of the channel biofilm clumps are pushed out of the domain.
Nevertheless, the qualitative feature of periodical removal of large chunks of biofilm rather than a continuous outflow of cells is well reproduced in agreement with the experimentally observed intermittency.

\begin{figure}[H]
    \centering
    \includegraphics[width=0.8\linewidth]{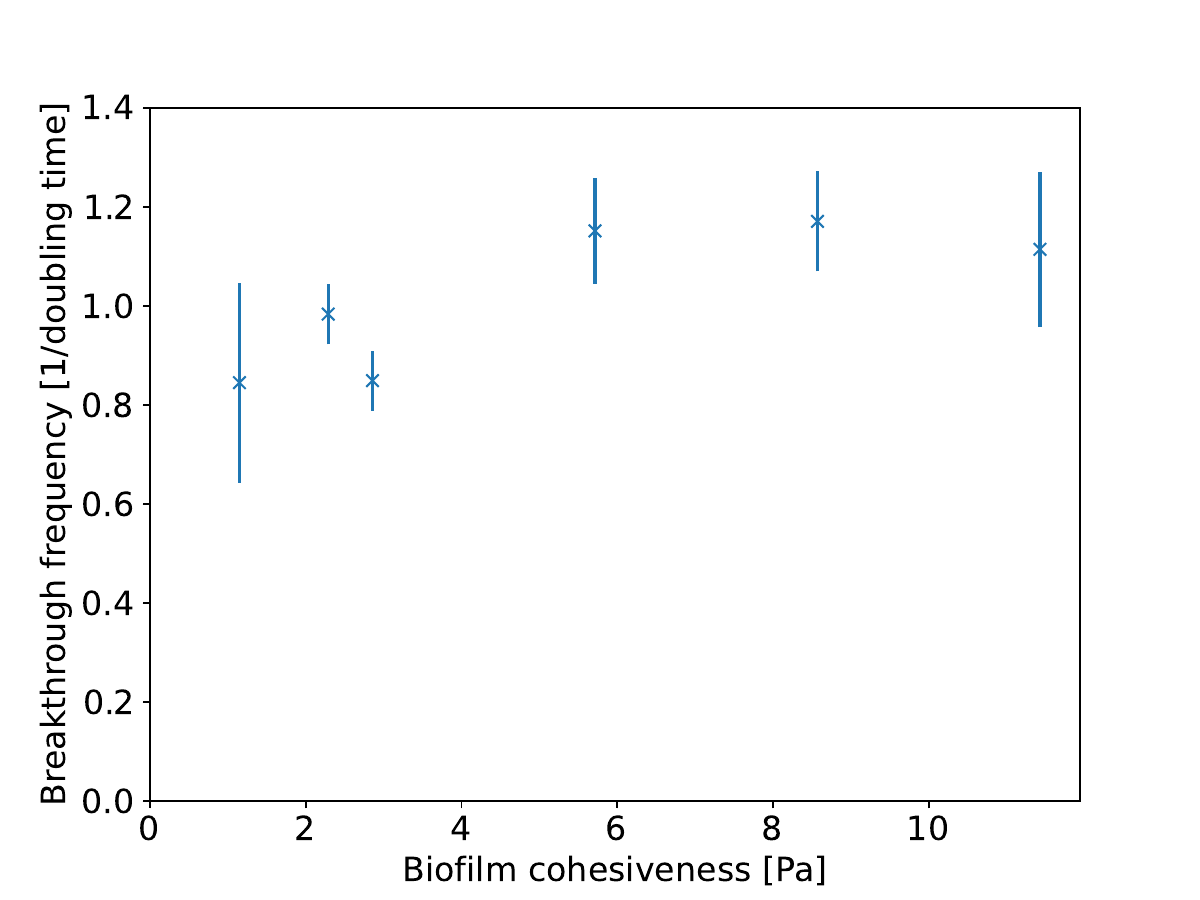}
    \caption{Frequency of detachments as a function of cohesiveness. Error bars show the standard error of the mean over 5 statistically independent simulations. The doubling time was implemented as $\tau_2$ = ln(2)/$r_{\text{growth}}$}
    \label{fig:sim_freq_vs_cohesion}
\end{figure}

The frequency of detachment events as a function of biofilm cohesion is shown in Figure~\ref{fig:sim_freq_vs_cohesion}.

Over the whole range of cohesiveness, there is very little variation of frequency.
This confirms the observation from the experiments that while cohesion plays a crucial role in determining the qualitative regime of biofilm intermittency, it has little influence on the frequency of detachment events.
The time-scale of intermittency is instead mostly set by the doubling time, which is an indirect measure of nutrient availability.

In the experiments, the control of shear stress and nutrient availability influenced the structural and rheological properties of the biofilm and leads to variations in the frequency of detachment events. 
In contrast, the model lacks variability of nutrients and assumes that they are always available. The model's range of biofilm cohesiveness is significantly larger than experimentally calculated values, yet resulted in only a small change in the frequency of detachment events, giving a strong indication that this parameter is not relevant in determining the frequency of detachment events.
In essence, the frequency of detachment events depends not only on biofilm cohesion but also on the interplay between biofilm cohesiveness, the shear stress applied to the biofilm, and nutrient availability.

The simulations only cover the regime of intermittent flow path formations; the other regimes were not computationally accessible.

\section{Conclusion}\label{sec:conclusion}

In this study, we conducted experiments in biofilm-affected microfluidic channels under various liquid flows. We investigated the dynamic balance between biofilm attachment and accumulation mechanisms, and the detachment mechanism due to shear forces induced by the flow, along with the resulting development of intermittent flow paths. We compared the results with a model based on coarse-grained molecular dynamics and Lattice Boltzmann hydrodynamics. Our findings revealed an intermittent regime where biofilm growth and flow path formation coexist in dynamic balance, governed by specific ranges of flow velocity, shear forces, nutrient availability, and biofilm cohesiveness, which all interact to define the frequency of biofilm detachment events.

\begin{itemize}
    \item The experiments demonstrate that intermittent flow path development is highly dependent on the liquid mean velocity, as the nutrient solution, while delivering essential nutrients to the biofilm, also generates shear forces. 
    \item While high Eu\textsubscript{co} at low Re number should favor biofilm growth over shear-induced detachment from a hydrodynamical point of view, it was observed that the very low mean velocity of nutrients in the E1.1. experiment effectively prevented growth due to nutrient transport limitation.
    We have observed that increasing the nutrient concentration in the injected solution tends to improve conditions for growth but did not investigate it in detail.
    \item When, indeed, at very low mean velocities, the concentration of nutrients were increased (E1.2. experiment), the biofilm could grow while still no penetrating flow paths developed, which we attributed to the high permeability of the biofilm itself. The biofilm adapted to the low-shear conditions by establishing itself as having a loose and porous structure.
    \item In cases where the mean velocity was high enough at a given nutrient concentration and thus the nutrient flux, biofilm develops. A dynamic balance was achieved between biofilm growth and shear stress if the velocity remained within a range that did not exceed the critical shear stress (as observed in the E2, E3, and E5 experiments). In this case, the biofilm withstands shear forces to a certain point until the shear forces overcome the biofilm's adhesive or cohesive forces, resulting in the dynamic growth and detachment pattern observed in E2, E3, and E5. This velocity range has been shown to correspond to specific Re and Eu\textsubscript{co} numbers.
    \item We added pore bodies to the flow channel, introducing regions of lower shear stress. We observed that the biofilm adapted to these low-shear conditions but was then unable to survive in the channel (E7 experiment), even though the flow and nutrient conditions were identical to those in the experiments without pore bodies. This provides evidence of the biofilm's adaptability to different shear conditions. 
    \item For increasing mean velocities of the nutrient flow where the Eu\textsubscript{co} is low, the frequency of detachment events and, as a result, the frequency of flow paths initially increase, then totally disrupts the dynamic balance, i.e. above the critical shear stress (E4 and E6 experiments). This results in higher detachment rates and inhibits biofilm growth.
    \item In agreement with the experimental observation, the comparison with a numerical Lattice-Boltzmann biofilm model confirmed the mechanisms of the intermittent regime of biofilm growth and the development of flow paths. The simulations varied biofilm cohesiveness, which influenced the formation of biofilm chunks and flow path development, consistent with experimental observations. However, changes in the frequency of flow paths were observed in the experiments when considering flow velocity and nutrient availability — factors which are not included in the model. Ultimately, the frequency of detachment events is governed by the interplay between biofilm cohesion, shear stress, and nutrient availability.
\end{itemize}

In summary, the study found experimental and numerical evidence of the existence of an intermittent flow-path regime in a dynamic balance with biofilm growth, which is defined within certain ranges of flow velocity and corresponding shear forces, nutrient availability, and biofilm cohesiveness.

\section*{Acknowledgements}
Funded by the Deutsche Forschungsgemeinschaft (DFG, German Research Foundation) under Project Number 327154368-SFB 1313 and 390740016 - EXC 2075 ``SimTech''. \newline
This work was supported by Michael Neubauer, Stephan Warnat, and Andrew Lingley and was performed in part at the Montana Nanotechnology Facility, a member of the National Nanotechnology Coordinated Infrastructure (NNCI), which is supported by the National Science Foundation (Grant ECCS-2025391). 

\section*{Data availability}
The complete source code including input data of the shown experiments that support the findings of this study is openly available in the Data Repository of the University of Stuttgart (DaRUS) at https://doi.org/10.18419/darus-4314.

\section*{Conflicts of interest}
There are no conflicts of interest to declare.

\section*{Author contributions}
K. Bozkurt: Conceptualization, Data curation, Investigation, Methodology, Writing - original draft.
C. Lohrmann: Conceptualization, Data curation, Investigation, Methodology, Software, Writing - original draft.
F. Weinhardt: Conceptualization, Investigation, Methodology, Writing - original draft.
D. Hanke: Data curation, Investigation, Writing - original draft.
R. Hopp: Data curation, Investigation, Writing - original draft.
R. Gerlach: Conceptualization, Writing – review and editing
C. Holm: Conceptualization, Writing - Review \& Editing, Supervision, Funding acquisition.
H. Class: Conceptualization, Writing - Review \& Editing, Supervision, Funding acquisition.

\section{Supplementary information}

\subsection{Introduction}

The supporting information details the experimental setup, provides videos of all experiments, the plots of the frequency of the detachment events, and the simulation parameters.

\subsection{The experimental setup}

\begin{figure}[H]
    \centering
\includegraphics[scale=0.4]{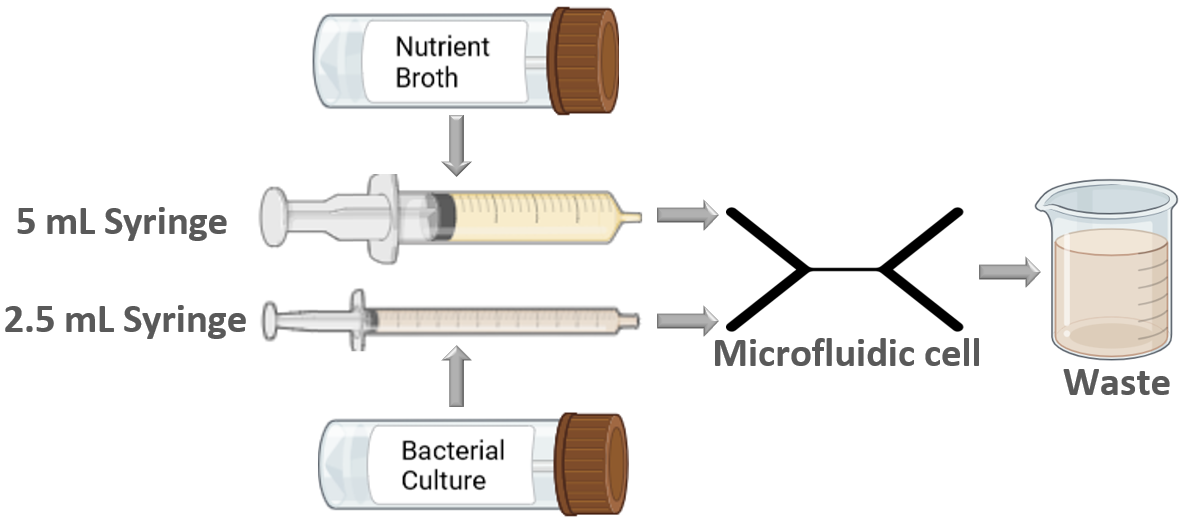}
    \caption{Experimental setup. It includes two syringes, one is for nutrient broth, one is for the bacterial culture. The microfluidic channels have two inlets and outlets. While the bacterial culture was injected through the bottom inlet, the nutrient broth was injected through the top inlet. The waste at the outlet was elevated to minimize the potential for gas bubble formation}
    \label{fig:Setup}
\end{figure}

The experimental setup (Figure \ref{fig:Setup}) consists of a 2.5 mL glass syringe for the bacterial liquid culture and a 5 mL glass syringe for the nutrient broth. The syringes are connected to the microfluidic channel via PTFE tubes with an inner diameter of 0.5 mm and T-valves, with the other outlet of the T-valves directed to a waste container. The bacterial liquid culture is injected through the bottom inlet, while the nutrient broth solution is injected through the top inlet.

\subsection{Amount of nutrients supplied to the microfluidic channel}

Table \ref{tab:nutrients} shows the amount of nutrients supplied to the microfluidic in all experiments. The concentration of nutrient broth solution is 8 g/L. For a better understanding of the experiment E1, E1.3. was performed with 16 g/L nutrient solution.

\begin{table}[h!]
\centering
\caption{Summary of experiments with provided nutrient broth flow rates. E1, E2, E3, and E4 were conducted in the narrow channel. E5 and E6 was conducted in the wide channel. E7 was conducted in the narrow channel with pores}
\begin{tabular}{cccc}
\toprule
\textbf{Experiments} & 
\textbf{\makecell{$Q_{\text{n.broth}}$ \\ (\SI{}{\micro\liter\per\second})}} & 
\textbf{\makecell{Provided \\ Nutrients \\ (\SI{}{\gram\per\minute})}} & 
\textbf{\makecell{Provided \\ Nutrients \\ (\SI{}{\micro\gram\per\minute})}} \\ 
\midrule
E1 & $5 \cdot 10^{-4}$ & $2.4 \cdot 10^{-7}$ & 0.24 \\ 
E1.3 & $5 \cdot 10^{-4}$ & $4.8 \cdot 10^{-7}$ & 0.48 \\ 
E2 & $1 \cdot 10^{-3}$ & $4.8 \cdot 10^{-7}$ & 0.48 \\ 
E3 & $3.5 \cdot 10^{-3}$ & $1.7 \cdot 10^{-6}$ & 1.7 \\ 
E4 & $7 \cdot 10^{-3}$ & $3.4 \cdot 10^{-6}$ & 3.4 \\ 
E5 & $7 \cdot 10^{-3}$ & $3.4 \cdot 10^{-6}$ & 3.4 \\ 
E6 & $1.4 \cdot 10^{-2}$ & $6.7 \cdot 10^{-6}$ & 6.7 \\ 
E7 & $3.5 \cdot 10^{-3}$ & $1.7 \cdot 10^{-6}$ & 1.7 \\ 
\bottomrule
\end{tabular}
\label{tab:nutrients}
\end{table}

\subsection{The videos of all experiments}

Each video is named according to the experimental numbers (e.g, V-E1.3.mp4 for experiment E1.3, V-E2 for E2, etc.). \\
In addition to these videos, the video\textit{ V-Rolling} demonstrates how some biofilms exit the channel as if they were rolling, due to their cohesive strength. It can also be observed that the biofilm does not move synchronously with the fluid flow. Some detached biofilm fragments are observed to reattach, while others exit the channel. Also, shear forces act on the biofilm's surroundings, causing small biofilm clusters to gradually detach from the larger rolling biofilm and as a result, the biofilm breaks off from the microfluidic system.

\subsection{The plots of velocity profile and shear stress of E3 along the height and width of the channel}

In this section, the velocity profile and shear stress of the E3 experiment along both height and width are described. Furthermore, the velocity profile in 2D is also given. As stated in Section 2.1.2, the flow profile in the channels has been derived based on \cite{Joseph1868}, because the channels are rectangular in shape. The gained velocity profiles can also be found in \cite{Esfahani}.

\subsection{The graphs obtained from image processing}

The graphs below have been generated through individual analyses of all the images from each experiment. These graphs were used for estimated biofilm volume and calculating the frequency of detachment events. The selected points, where large biofilm fragments begin to accumulate immediately after detachment, are also indicated. Since no clear biofilm accumulation was observed, image processing could not be performed for the E1, E6, and E7 experiments.

\begin{figure}[H]
    \centering
    \includegraphics[scale=0.22]{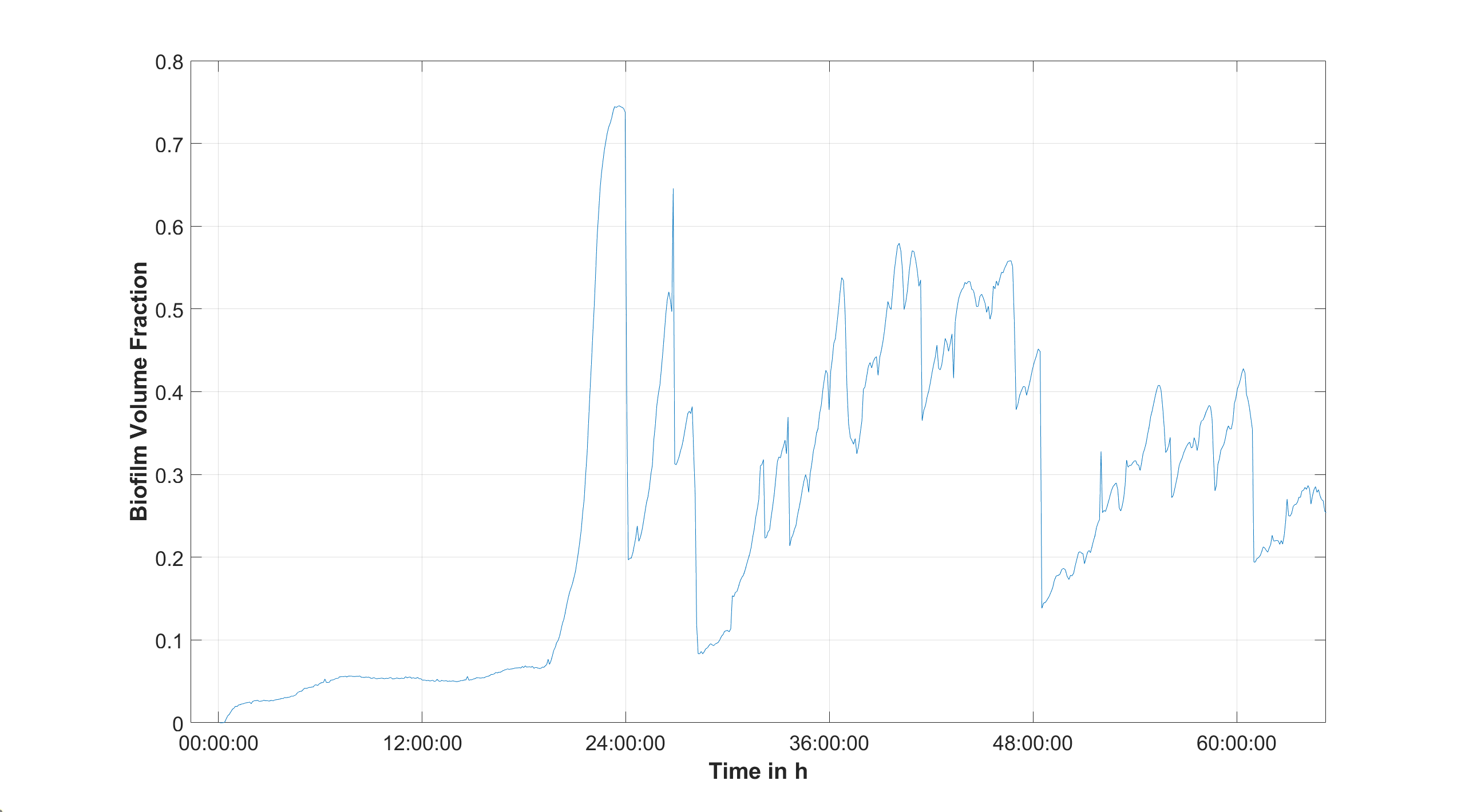}
    \caption{E2 ($v_0 = 0.23 \, \text{mm/s}$ in narrow channel). There was a dynamic balance with the attachment, biofilm accumulation and the detachment. }
    \label{fig:E2_markers}
\end{figure}

\begin{figure}[H]
    \centering
    \includegraphics[scale=0.22]{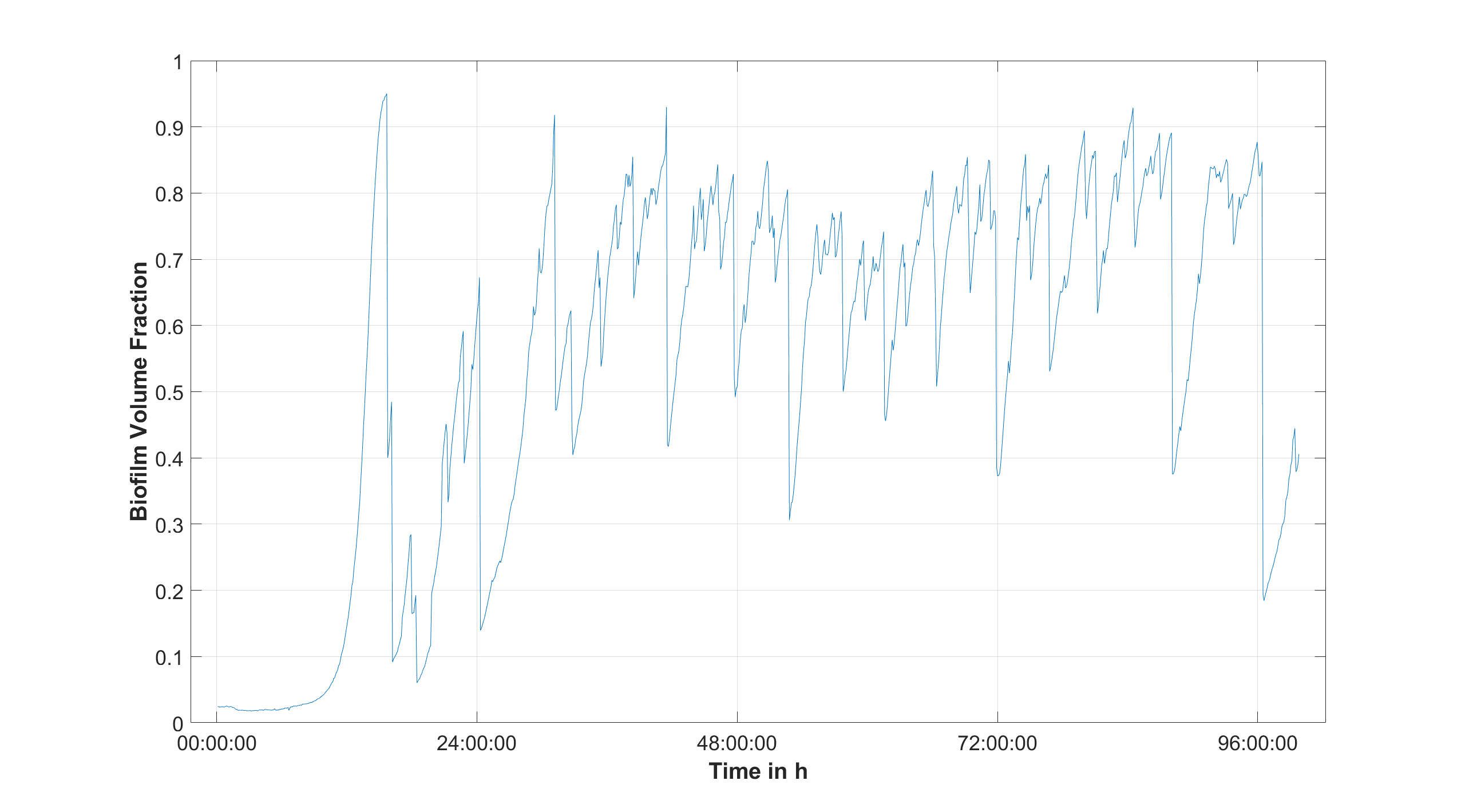}
    \caption{E3 ($v_0 = 0.8 \, \text{mm/s}$ in narrow channel). There was a dynamic balance with the attachment, biofilm accumulation and the detachment. This plot is the full length plot of the experimnet E3.1. }
    \label{fig:E3_womarkers}
\end{figure}

\begin{figure}[H]
    \centering
    \includegraphics[scale=0.22]{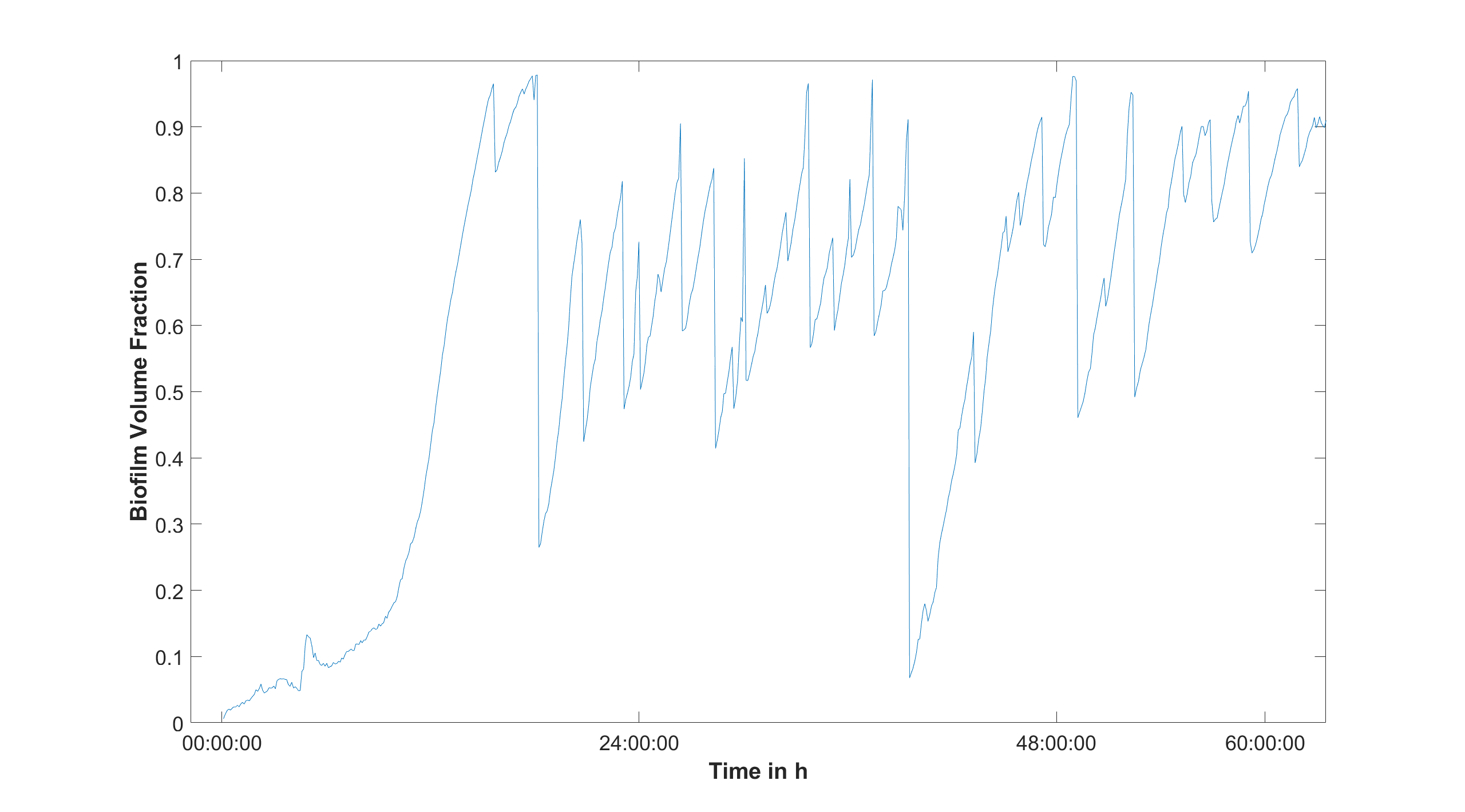}
    \caption{E3.2 ($v_0 = 0.8 \, \text{mm/s}$ in narrow channel). It was the second trial of the E3. There was a dynamic balance with the attachment, biofilm accumulation and the detachment. }
    \label{fig:E3.2_markers}
\end{figure}

\begin{figure}[H]
    \centering
    \includegraphics[scale=0.22]{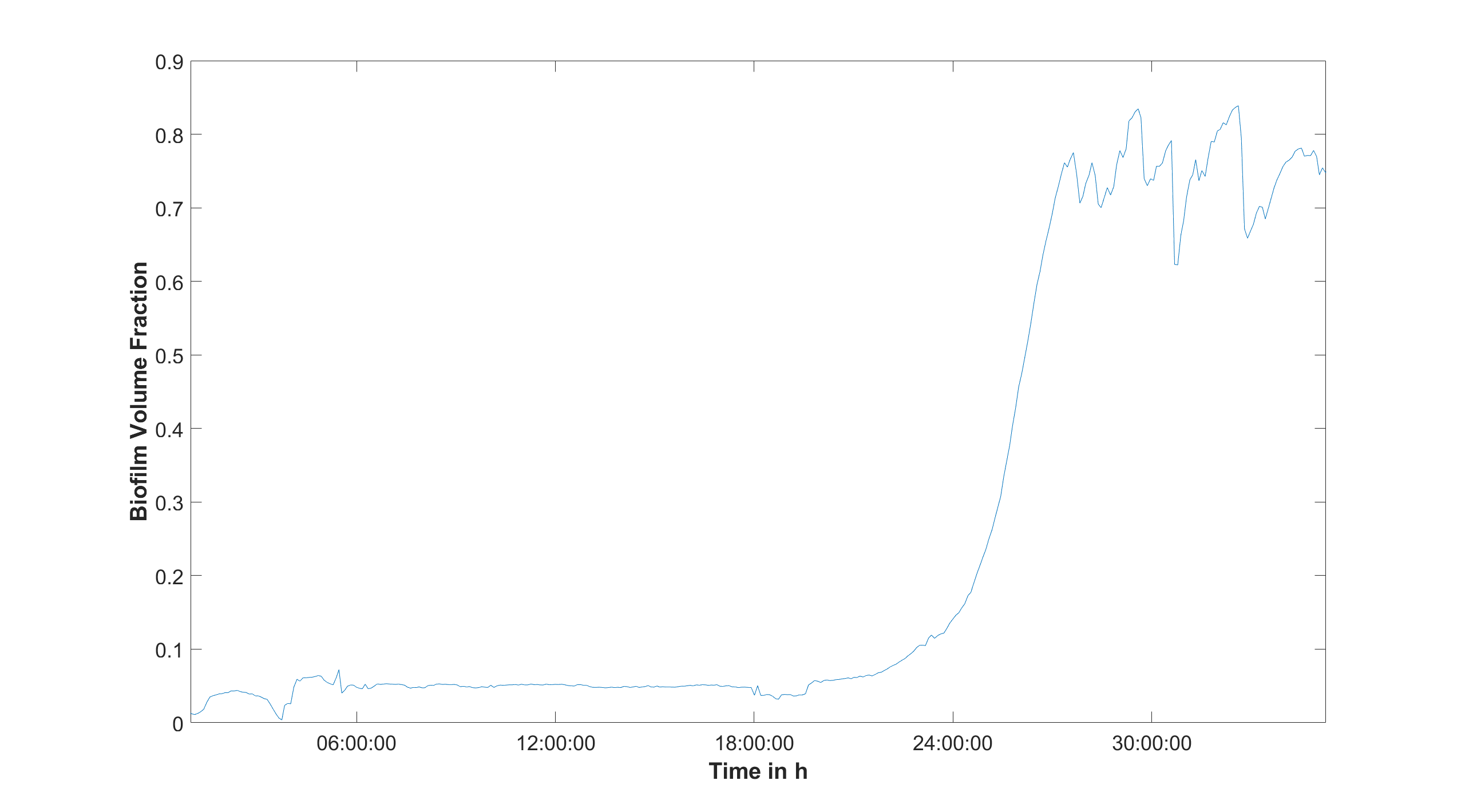}
    \caption{E4.1 ($v_0 = 1.6 , \text{mm/s}$ in narrow channel) was the first trial of E4. Initially, there was a dynamic balance between attachment, biofilm accumulation, and detachment. However, this balance was disrupted after approximately 34 hours, and biofilm accumulation could no longer be observed after that point.}
    \label{fig:E4.1_markers}
\end{figure}

\begin{figure}[H]
    \centering
    \includegraphics[scale=0.22]{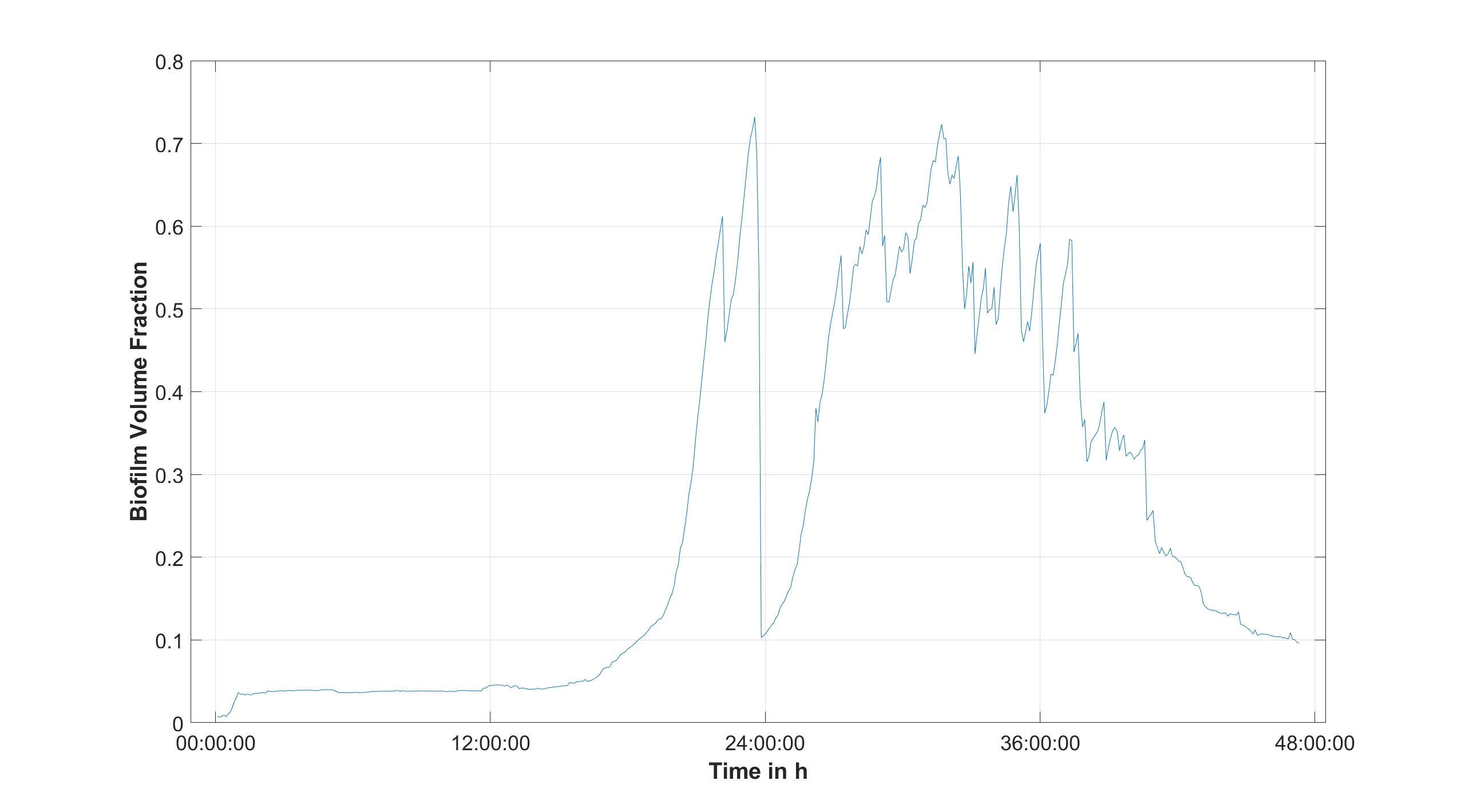}
    \caption{E4.2 ($v_0 = 1.6 , \text{mm/s}$ in narrow channel) was the second trial of E4. Initially, there was a dynamic balance between attachment, biofilm accumulation, and detachment. However, this balance was disrupted after approximately 35 hours, and biofilm accumulation could no longer be observed after that point.}
    \label{fig:E4.2_markers}
\end{figure}

\begin{figure}[H]
    \centering
    \includegraphics[scale=0.22]{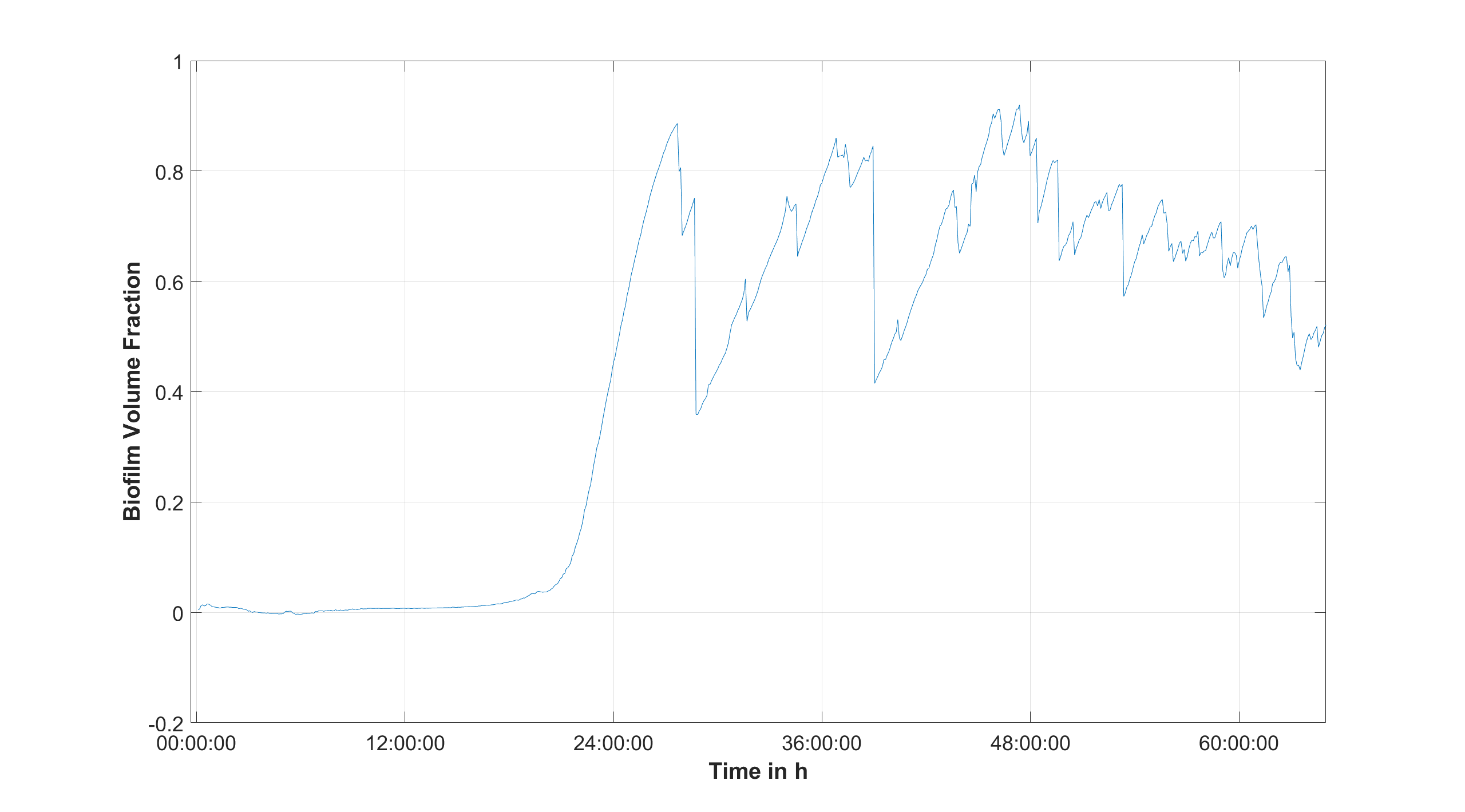}
    \caption{E5 ($v_0 = 0.8 \, \text{mm/s}$ in wide channel). There was a dynamic balance with the attachment, biofilm accumulation and the detachment.}
    \label{fig:E5.2_markers}
\end{figure}

\subsection{Simulation domain and parameters}
The parameters for the numerical simulations are listed and explained in Table \ref{tab:sim_params}.
For a more detailed explanation of the model implementation, we refer to the description in Ref.~\cite{lohrmann23a}.

\begin{table}[ht]
    \centering
        \caption{Parameters for the numerical simulations.}
    \begin{tabular}{l|l|l}
         symbol & explanation & value \\
         \hline
         $l_x$ & channel length & \SI{150}{\micro\meter}\\
         $l_y$& channel width & \SI{50}{\micro\meter}\\
         $l_z$& channel height & \SI{35}{\micro\meter}\\
         $\Delta x$& LB grid spacing & \SI{1}{\micro\meter} \\ 
         $\Delta t_\text{LB}$ & LB time step & \SI{0.03}{\second}\\
         $\mu$  & fluid dynamic viscosity & \SI{2e-9}{\pascal\second}\\
         $\rho_\text{fluid}$ & fluid mass density & \SI{1000}{\kilo\gram\per\meter\cubed} \\
         $\Delta t_\text{MD}$ & particle simulation time step & \SI{0.01}{\second} \\
         $\Delta t_\text{model}$ & bacterial model time step & \SI{3}{\second} \\
         $N_\text{beads}$   & Number of spheres per cell & 5 \\
         $\rho_\text{bact}$   & cell mass density & \SI{1000}{\kilo\gram\per\meter\cubed} \\
         $r_\text{bact}$   & cell radius & \SI{1}{\micro\meter} \\
         $l_\text{bact}$  & cell maximum length & $10 \, r_\text{bact}$\\
         $\tau_2$  & doubling time & \SI{120}{\minute} \\
         $\Delta l/l$  & daughter cell length variation & 0.05 \\
         $\bareFrict$ & friction coefficient & \SI{1.7e-8}{\micro\meter\pascal\second}\\
         $\epsilon^\text{LJ}$   & cell-cell interaction strength & is varied\\
         $r_\text{cut}^\text{LJ}$   & cell-cell interaction range & $3 \,r_\text{bact}$\\
         $\epsilon^\text{WCA}$   & cell-surface interaction strength & $\epsilon^\text{LJ}$ \\
         $N^\text{anchor}$     & max. number of anchors per cell & 2 \\
         $r_\text{attach}$     & surface attachment range & $1.25 \, r_\text{bact}$\\
         $r_\text{bond}$     & surface bond length &$1.25 \, r_\text{bact}$ \\
         $k_\text{bond}$     & surface bond strength & \SI{1.2e-6}{\pico\newton\per\micro\meter} \\
         $r_\text{detach}$     & surface detachment range &$ 2 \, r_\text{bact}$ \\
              & & \\
    \end{tabular}

    \label{tab:sim_params}
\end{table}

% \begin{appendices}
% \section{Section title of first appendix}\label{sec:appendix}
% \end{appendices}

\end{document}